\documentclass[aps,prl,floatfix,superscriptaddress,preprintnumbers,nofootinbib,twocolumn,longbibliography]{revtex4-2}
\pdfoutput=1

\usepackage{amsmath, amssymb, amsfonts, graphicx, mathrsfs, verbatim, float, slashed, bbm, color, xcolor, xspace, enumerate, url, bbding, multirow, outlines,upgreek}
\usepackage[english]{babel}
\usepackage[colorlinks=true,linkcolor=foundationteal,citecolor=foundationteal]{hyperref}
\definecolor{foundationteal}{rgb}{0.3,0.7, 0.8}

\begin{document}

\title{Nuclear Fusion Inside Dark Matter}

\author{Javier F. Acevedo}
\affiliation{The McDonald Institute and Department of Physics, Engineering Physics, and Astronomy, Queen's University, Kingston, Ontario, K7L 2S8, Canada}

\author{Joseph Bramante}
\affiliation{The McDonald Institute and Department of Physics, Engineering Physics, and Astronomy, Queen's University, Kingston, Ontario, K7L 2S8, Canada}
\affiliation{Perimeter Institute for Theoretical Physics, Waterloo, Ontario, N2L 2Y5, Canada}

\author{Alan Goodman}
\affiliation{The McDonald Institute and Department of Physics, Engineering Physics, and Astronomy, Queen's University, Kingston, Ontario, K7L 2S8, Canada}

\begin{abstract}
A new dynamic is identified between dark matter and nuclei. Nuclei accelerated to MeV energies by the internal potential of composite dark matter can undergo nuclear fusion. This effect arises in simple models of composite dark matter made of heavy fermions bound by a light scalar field. Cosmologies and detection prospects are explored for composites that catalyze nuclear reactions in underground detectors and stars, including bremsstrahlung radiation from nuclei scattering against electrons in hot plasma formed in the composite interior. If discovered and collected, this kind of composite dark matter could in principle serve as a ready-made, compact nuclear fusion generator.
\end{abstract}

\maketitle

\section{Introduction}
The presence of dark matter has become manifest through galactic dynamics, the lensing of light, and temperature fluctuations in the cosmic microwave background. But setting aside these gravitational signifiers, little is known about dark matter despite extensive laboratory and astrophysical efforts. It is a high priority of modern science to uncover dark matter, identify its mass and couplings, and determine what influence it may have on other particles that compose the known universe. 

In the past decade theorists have enunciated how a certain variety of dark matter could bear a striking resemblance to known matter. Atoms, nuclei, and nucleons, which comprise the bulk of known particles, are all built from fundamental fermions -- electrons, protons, quarks -- bound together by photons and gluons into composite states. Similarly, dark matter could also be comprised of many particles bound together in a composite state \cite{Nussinov:1985xr,Bagnasco:1993st,Alves:2009nf,Kribs:2009fy,Lee:2013bua,Krnjaic:2014xza,Detmold:2014qqa,Jacobs:2014yca,Bramante:2018tos,Ibe:2018juk,Coskuner:2018are,Bai:2018dxf,Bai:2019ogh,Bramante:2019yss}. One simple composite dark matter model consists of fermions ($X$) bound by a new attractive force provided by a massive scalar field ($\varphi$) \cite{Wise:2014jva,Wise:2014ola,Hardy:2014mqa,Hardy:2015boa,Gresham:2017zqi,Gresham:2017cvl,Gresham:2018anj}. If this force is strong enough, then in the early universe large dark matter states would be built from successive fusion of $X$ particles into increasingly massive states, in a process similar to big bang nucleosynthesis (BBN). In the absence of the repulsive Coulomb force between protons in Standard Model nuclei, these dark composites can become extremely massive after accumulating oodles of $X$ particles. As we will see in this work, if $X$ has a TeV-EeV mass, this can imply dark matter composite masses ranging from a few micrograms to thousands of tons.

\begin{figure}[!h]
    \includegraphics[width=0.34\textwidth]{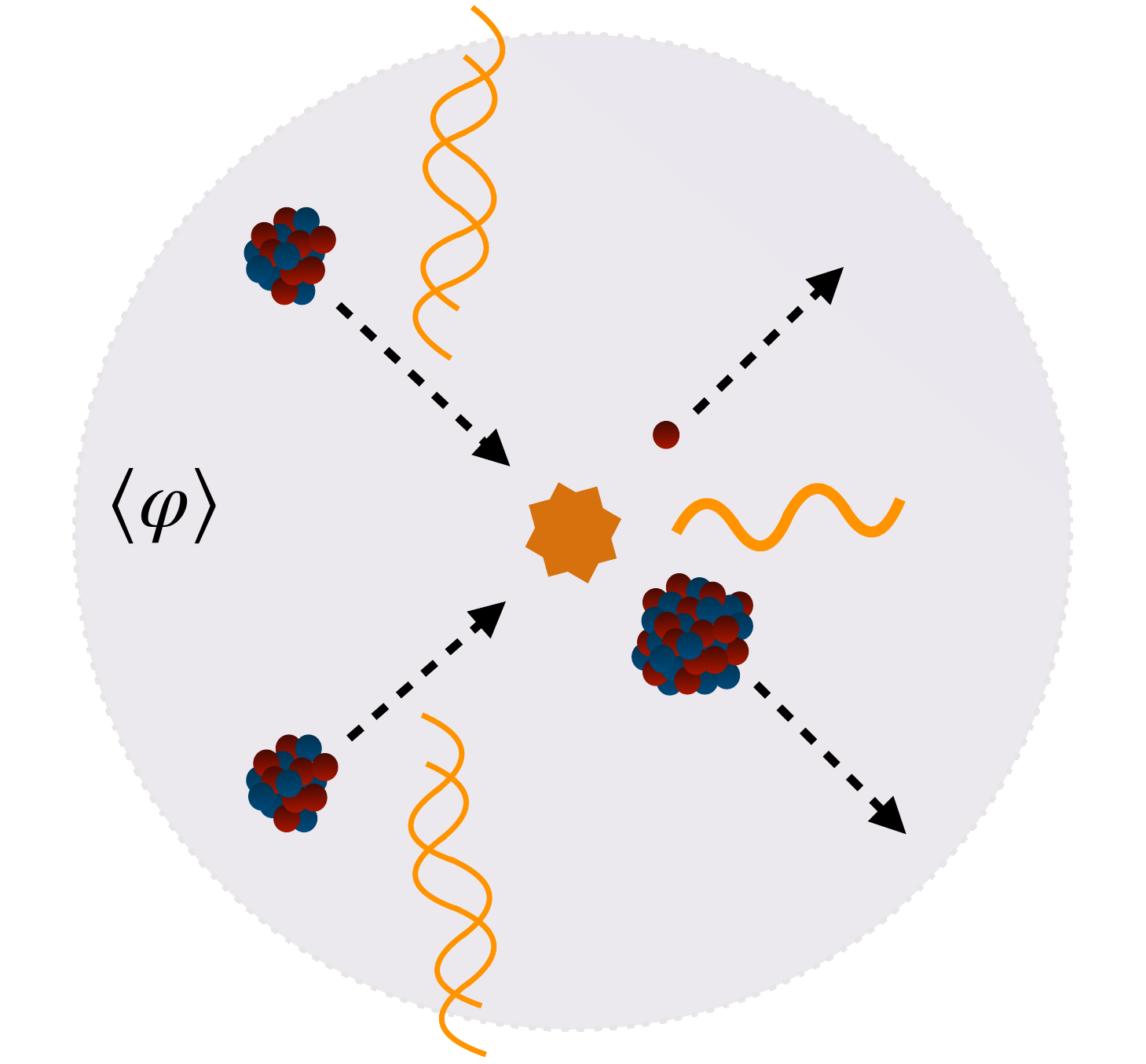}
    \caption{Schematic of nuclei accelerated by the potential $\left\langle \varphi \right\rangle$ inside a dark matter composite, resulting in ionization, bremsstrahlung radiation, and thermonuclear fusion.} 
    \label{fig:dfus}
\end{figure}

We have found that such large dark matter (DM) composites imply novel dynamical interactions with Standard Model nuclei. In this paper we present these newly identified dynamics. Large composite dark matter can cause Standard Model (SM) nuclei to accelerate, radiate, and fuse in the composite interior, as shown in Fig.~\ref{fig:dfus}. These dynamics occur because the scalar field binding $X$ particles together can have an extremely high potential $\left\langle \varphi \right\rangle$ inside the dark matter composite. Under the influence of this potential, SM nuclei are accelerated to energies $\Delta E \sim g_n \left\langle \varphi \right\rangle \sim {\rm MeV}$, sufficient to initiate nuclear fusion and radiation from high energy collisions, even for a miniscule Yukawa coupling $g_n$ between $\varphi$ and nucleons. This implies new signatures and even potential uses for composite dark matter, including nuclear fusion and bremsstrahlung radiation as unique signatures in particle detectors, nuclear reactions in stars and planets, and speculatively the use of composites as compact fusion reactors. In addition, we find that the largest fusion-capable composites would make white dwarfs explode.

\section{Heavy Composite Cosmology}
We begin with the cosmology of very large composites, formed from heavy asymmetric fermions with masses ranging up to an EeV. As we will see, dark composites formed from such heavy fermions can have large internal potentials that accelerate Standard Model nuclei to fusion temperatures. The cosmology of up to $10^{10}$ GeV mass asymmetric dark matter, motivated by high scale baryogenesis mechanisms like Affleck-Dine \cite{Affleck:1984fy,Dine:1995kz}, has been detailed in \cite{Bramante:2017obj}. In asymmetric dark matter models \cite{Zurek:2013wia,Petraki:2013wwa}, an initial dark sector particle asymmetry sets the dark matter relic abundance, and typically dark matter freeze-out annihilation eliminates most of the symmetric dark matter abundance, $i.e.$ $X + \bar X \rightarrow$ SM, leaving behind a residual asymmetric abundance of $X$ particles. For a heavy asymmetric dark matter scenario \cite{Bramante:2017obj}, the abundance of dark fermions is subsequently depleted (along with the baryon abundance) by $e.g.$ the decay of a field some time after freeze-out. The amount the asymmetric dark matter (and baryon) abundance is depleted by this decay is given by $\Omega_{DM}^{dep} = \Omega_{DM} \zeta$, where $\zeta = s_{before}/s_{after} $ is the ratio of entropy density in the universe before and after the field decays \cite{Bramante:2017obj}. In the models that follow, we will assume that dark fermions freeze-out to an initial abundance that is later diluted through the decay of a metastable field \cite{Dine:1995kz,Banks:1993en,Randall:2015xza,Berlin:2016vnh,Bernal:2019mhf,Evans:2019jcs}, or a phase transition/second phase of inflation \cite{Burgess:2005sb,Wainwright:2009mq,Davoudiasl:2015vba,Hoof:2017ibo,Breitbach:2018ddu,Hambye:2018qjv}. This means that right after its freeze-out, dark matter's abundance will be larger by a factor of $ \zeta^{-1}$, relative to a cosmology without subsequent depletion. We will see that this relative overabundance of $X$ after freeze-out leads to the formation of rather large DM composites.

Asymmetric DM composites made of sub-TeV mass fermions have been studied at length in \cite{Wise:2014jva,Wise:2014ola,Hardy:2014mqa,Hardy:2015boa,Gresham:2017zqi,Gresham:2017cvl,Gresham:2018anj}. Here we consider heavier fermions. The Lagrangian,
\begin{align}
    \mathcal{L} = \frac{1}{2}(\partial \varphi)^{2} &+ \bar{X}(i\gamma^{\mu}\partial_{\mu}-m_{X})X  +g_{X}\bar{X}\varphi X-\frac{1}{2}m_{\varphi}^{2}\varphi^{2} \nonumber \\&
    +g_n \bar n \varphi n + \mathcal{L}_{SM},
\label{eq:lag}
\end{align}
includes the scalar $\varphi$ which provides an attractive force that binds together $X$ fermions. The second to last term couples $\varphi$ to SM nucleons $n$, where this is the simplest renormalizable coupling to SM particles. Once enough $X$ particles are bound together, fermionic composites will reach a saturation point after the composite radius exceeds $R_X \gtrsim m_{\varphi}^{-1} $, at which point the interior density becomes approximately constant, $\rho_{c} = \bar m_X^4/ 3 \pi^2 $, where $\bar m_X$ is the constituent mass, $i.e.$ the effective mass of $X$ inside the composite \cite{Gresham:2017zqi,Gresham:2017cvl}. For the cosmological formation we consider hereafter, composites are saturated well before they finish forming, $cf.$ Eq.~\eqref{eq:Nc}. The constituent mass for a saturated composite is given by \mbox{$\bar m_X \simeq m_X - E_{X}$}, where $E_X$ is the binding energy per $X$. In terms of bare masses and couplings, in a saturated composite
\mbox{$
\bar m_X \simeq \left[3 \pi m_X^2 m_\varphi^2/(2 \alpha_X) \right]^{1/4},
$}
where the $\varphi-X$ coupling constant is $\alpha_X \equiv g_X^2/4 \pi$. For parameters we consider, the binding energy is close to the unbound $X$ mass, $E_X \sim m_X$, and so $\bar m_X \ll m_X$. This means the composite state of $X$ particles has a total mass $M_X \equiv N \bar m_X$, which is much less than the mass of unbound $X$ particles, $N m_X$. As a consequence, after the composite is assembled, the mass density of dark matter in the universe decreases by a factor $\bar m_X /m_X$, where the mass loss is accounted for by the emission of $\varphi$ radiation. 

Fermion composites will begin to assemble in the early universe by forming two-fermion bound states, where binding will occur so long as $\alpha_X^2 m_X \gtrsim m_\varphi$ and \mbox{$\alpha_X \gtrsim 0.3 \left(m_X/10^7\,{\rm GeV} \right)^{2/5} \left(\zeta/10^{-6} \right)^{1/5}$} \cite{Wise:2014jva}. After two-fermion states form, composites will build up through processes like $X_N + X_N \rightarrow X_{2N} + \varphi$, where $X_N$ is a bound state formed from $N$ fermions. At the temperature of composite assembly $T_{ca}$, an estimate for the number of constituent particles in a typical composite can be obtained by comparing the $X_N$ interaction rate to the Hubble rate \cite{Hardy:2014mqa,Gresham:2017cvl}, $n_{X_N} \sigma_{X_N} v_{X_N} / H \sim 1$. Re-expressing this in terms of the $X$ number density $n_X = n_{X_N}/N$, the $X$ composite cross-section $\sigma_{X} = \sigma_{X_N}/N^{2/3}$ (where $R_X$ scales as $N^{1/3}$ in the saturation regime), and the $X$ velocity $v_{X_N} = v_{X}/N^{1/2}$, we arrive at an expression for the number of $X$ particles in a typical composite,
\begin{align}
    \label{eq:Nc}
    &N_{c}  = \left(\frac{2 n_X \sigma_X v_X}{3H} \right)^{6/5} = \left(\frac{20 \sqrt{g^*_{ca}} T_{r} T_{ca}^{3/2} M_{pl}}{\bar m_X^{7/2} \zeta} \right)^{6/5} \\
    &\simeq 10^{27} \left(\frac{g^*_{ca}}{10^2} \right)^{3/5} \left(\frac{T_{ca}}{10^5\,{\rm GeV}} \right)^{9/5}  \left(\frac{5\,{\rm GeV}}{\bar m_X} \right)^{21/5}\left(\frac{10^{-6}}{\zeta} \right)^{6/5},\nonumber
\end{align}
where in the first equality we have included a factor of $2/3$ appropriate for composite assembly in a radiation-dominated universe \cite{Gresham:2017cvl}, in the second equality we have used a composite cross-section \mbox{$\sigma_X = 4 \pi R_c^2$ with $R_c \equiv (3 \bar m_X/4 \pi \rho_c)^{1/3}= (9 \pi/4)^{1/3}/\bar m_X$}, a velocity $v_X = \sqrt{T/\bar m_X}$, the Friedmann relation is \mbox{$3H^2 M_{pl}^2 = g^* \pi^2 T^4/30$} for Planck mass $M_{pl}$ and temperature $T$, and we estimate the $X$ density at the time of composite assembly as \mbox{$n_X = g_{ca}^* \pi^2 T_{ca}^3 T_r/ (30 \zeta \bar m_X)$}, where $T_r \simeq 0.8~{\rm eV}$ is the temperature at matter-radiation equality. For $m_X \gg \bar m_X$, the binding energy of these composites is $E_X \sim m_X$, and composite assembly will finish around the temperature of $X$ freeze-out, $T_{ca} \sim m_X/10$.

A few more facets of heavy composite cosmology are worth emphasizing. First, because the constituent mass $\bar m_X \ll m_X$ determines the final density of DM, heavy asymmetric DM composites can account for the baryon-DM density coincidence: the present-day DM density approximately matches the baryon density, $\Omega_{DM} / \Omega_{B} \sim 5$. For asymmetric DM, this coincidence can be explained by having a single particle asymmetry that determines both the baryon and DM relic abundances. While a naive prediction for the constituent DM mass relative to the baryon mass is then $\bar m_X \sim 5 m_b \sim 5 ~{\rm GeV}$, in the case that heavy asymmetric DM freezes out before the electroweak phase transition, electroweak sphalerons can dilute baryon number, leading to a looser prediction $\bar m_X \sim 1-1000$ GeV \cite{Zurek:2013wia}. Second, as previously discussed, in an asymmetric DM cosmology the symmetric DM component ($X  \bar X$) is depleted via annihilation. In the case of the heavy DM detailed above, $X \bar X$ annihilation to $\varphi$ will deplete $X \bar X$ to a sub-DM relic density, if the annihilation cross-section $\sigma_a v \simeq 3\pi \alpha_X^2/(8m_X^2) \gtrsim 10^{-36} {\rm cm^2} \times \zeta $  \cite{Bramante:2017obj}, which corresponds to 
$
  \alpha_X \gtrsim 0.3  \left(m_X/10^7\,{\rm GeV} \right) \left(\zeta/10^{-6} \right)^{1/2},$
although this restriction weakens if $X\bar X$ are depleted by additional annihilation channels or other mechanisms. 

\section{Nuclear Acceleration, Radiation, and Fusion In Composite DM}

Substantial energy can be released by nuclei accelerated inside large DM composites, both through fusion and bremsstrahlung processes. With the structure and cosmology of heavy composites previously laid out, we now turn to nuclear acceleration, radiation, and fusion inside large composites. We begin with the potential inside a saturated composite $\langle \varphi \rangle \simeq \frac{m_X}{g_X}$, obtained by requiring the composite's internal potential \eqref{eq:lag} is minimized at equilibrium. Boundary conditions require that outside the composite the potential decays as
\begin{equation}
    \varphi(r)= \langle \varphi \rangle e^{-m_\varphi(r-R_X )} \left(\frac{R_X}{r}\right).
    \label{eq:phi}
\end{equation}

{\em Acceleration.} Nuclei with $A$ nucleons will have their momentum $p$ boosted to $p'$ as they enter the composite, according to \mbox{$p^2 + m_N^2 = p'^2 + (m_N-V_n)^2$}, where \mbox{$V_n=A g_n \langle \varphi \rangle = A g_n m_X /g_X$}. In the limit \mbox{$V_n \ll m_N$}, the second term can be expanded yielding \mbox{$p'^{2}-p^{2}=2m_{N}V_n$}. Nuclei will accelerate over a time determined by the field gradient at the composite boundary and the velocity $v_X$ at which the composite moves, $cf.$ Eq.~\eqref{eq:phi}, \mbox{$\tau_{\rm accel}\simeq (m_\varphi v_X)^{-1} (1+2V_n/m_N v_X^2)^{-1/2}$}.

\begin{figure*}[!th]
    \centering
     \includegraphics[width=0.495\linewidth]{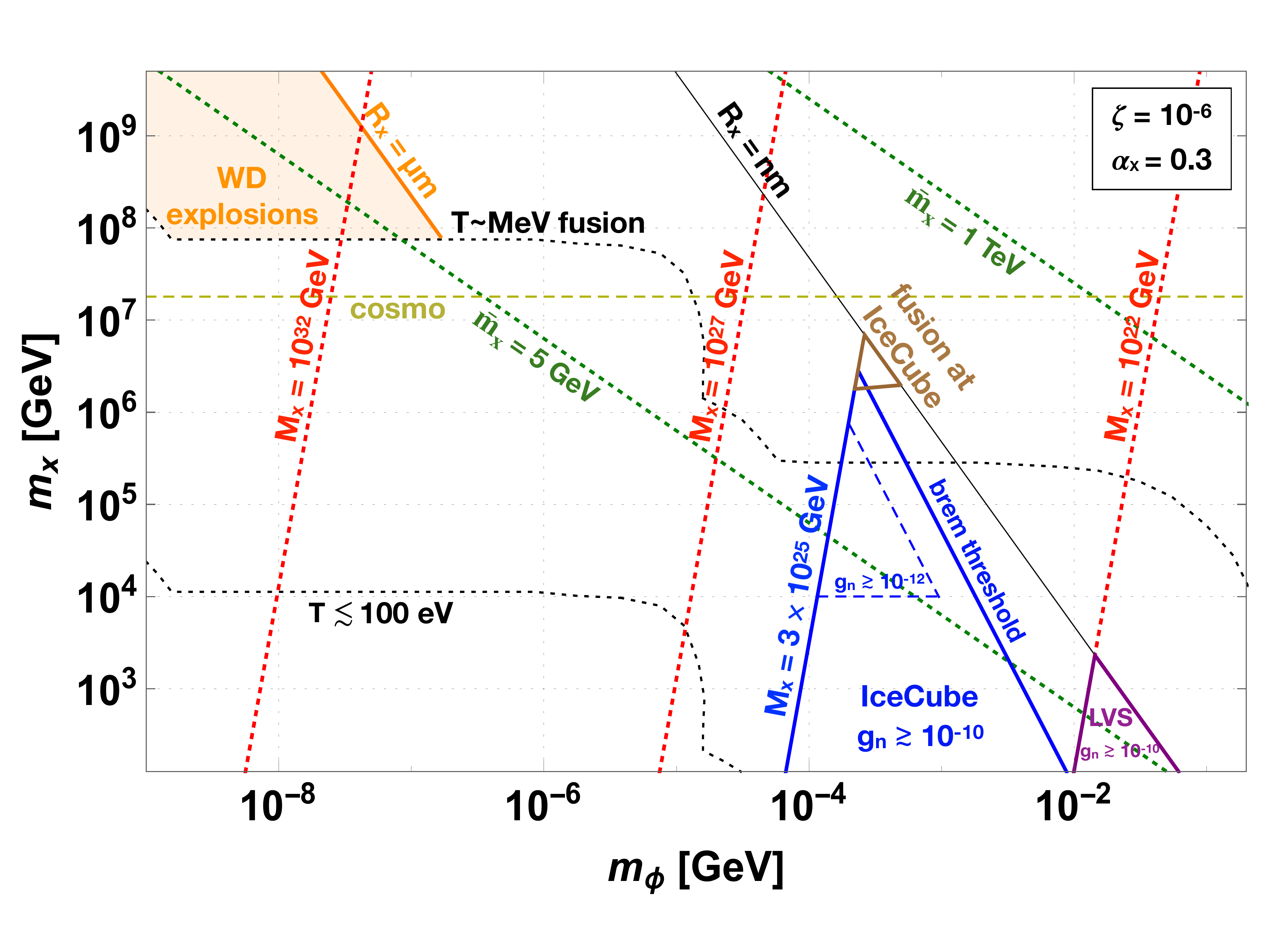}
     \includegraphics[width=0.495\linewidth]{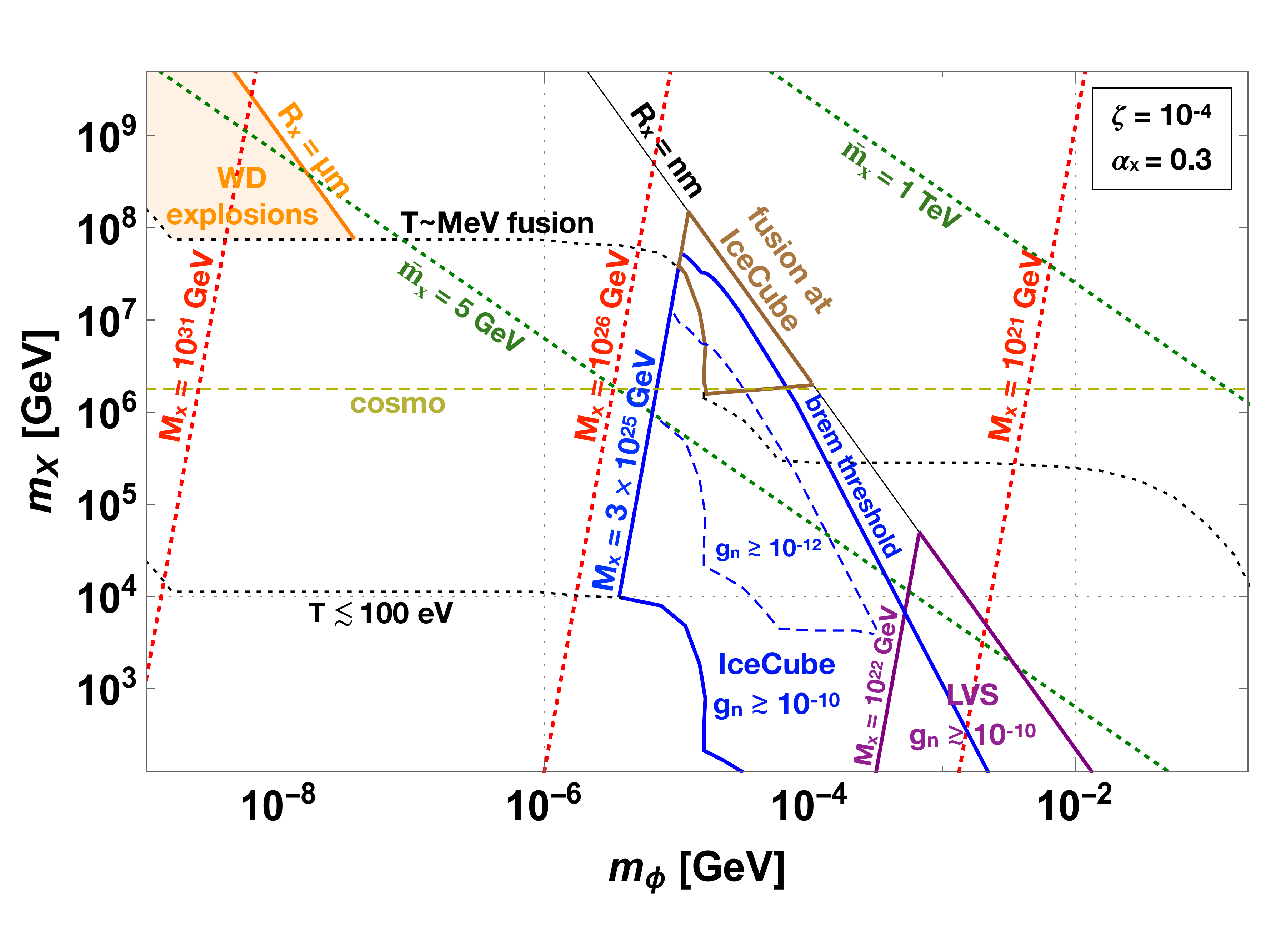}
     \includegraphics[width=0.495\linewidth]{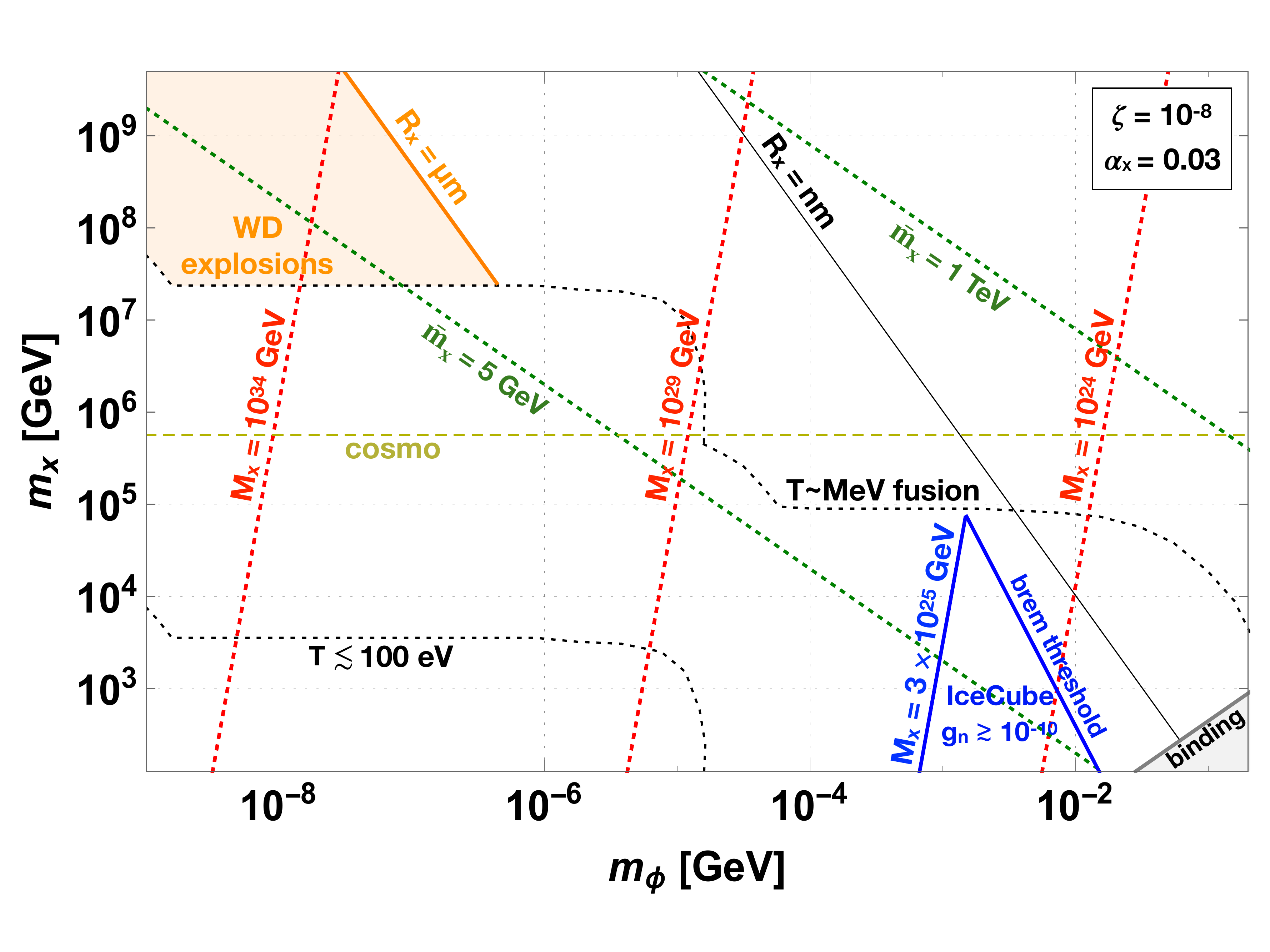}
     \includegraphics[width=0.495\linewidth]{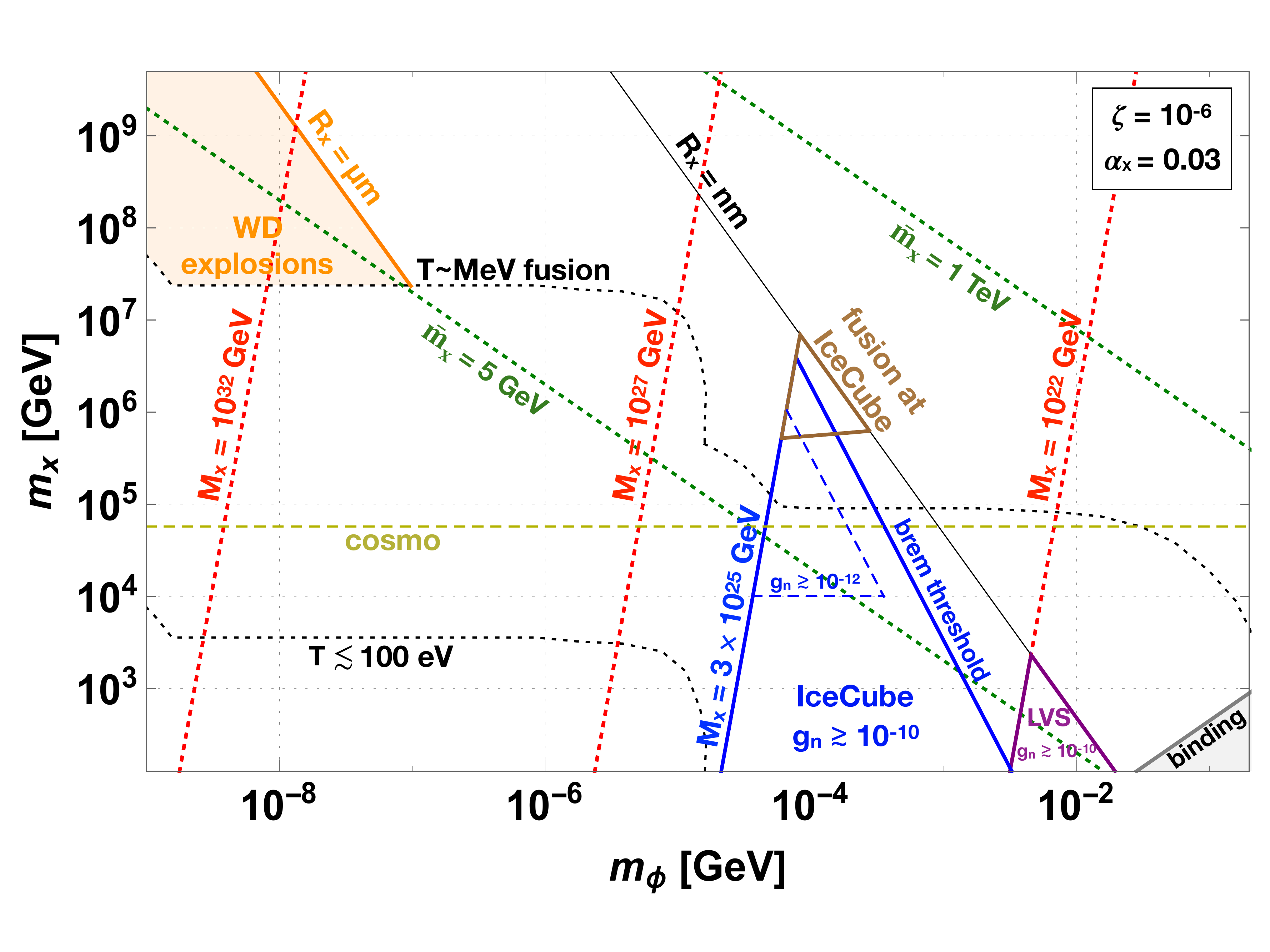}
    \caption{Heavy asymmetric composites that cause nuclei to radiate and fuse in their interiors, for fermion mass $m_X$, scalar mass $m_\varphi$, and $\varphi-X$ coupling $\alpha_X$. The total mass of the composites $M_X = N_c \bar m_X$ is shown with red dashed lines, determined by composite assembly after DM freeze-out, followed by a process that dilutes relic abundances by $\zeta$, $cf.$ Eq.~\eqref{eq:Nc}. The baryon and DM densities could arise from a common asymmetry for $\bar m_X \approx 5-1000$ GeV. Blue and purple regions show what composites can be discovered via bremsstrahlung radiation from ionized matter at IceCube and large volume scintillator (LVS) experiments, for $\varphi$-nucleon couplings $g_n \geq 10^{-10}$ and $g_n \geq 10^{-12}$ as indicated. For detection we require $R_X \gtrsim$ nm, so composites contain $\gtrsim 10$ atoms at solid Earth densities. Above the dotted line marked ``T$\sim$MeV fusion'', the max $g_n$ allowed by stellar bounds \cite{Hardy:2016kme,Knapen:2017xzo}, permits nuclei to be accelerated to MeV temperatures (the $T < 100 \ \rm eV$ line is similarly obtained). In the tan wedges, DM can cause nuclei to fuse at IceCube. The heaviest fusion-capable composites can be excluded by old $1.1-1.4 \ M_{\odot}$ white dwarfs not exploding, corresponding to central densities $\rho_* \sim 10^{8} - 10^{10} \ \rm g \ cm^{-3}$. Below the yellow dashed line marked ``cosmo,'' the composite cosmology detailed in the text is satisfied for $\alpha_X,\zeta$ values. A strong binding condition $\alpha_X^2 m_X \gtrsim m_\varphi$ \cite{Wise:2014ola} limits the bottom panels.}
    \label{fig:doxy}
\end{figure*}

{\em Ionization.} For parameters in Fig.~\ref{fig:doxy} saturated composites crossing terrestrial material at speeds $v_X \simeq 10^{-3}$ will accelerate nuclei on a timescale $\tau_{\rm accel}\lesssim 10^{-18} \ \rm s$ due to the sharp gradient of the potential. This timescale is shorter than both the electron orbital period $(10 \ {\rm eV})^{-1}\simeq 10^{-17} \ \rm s$ and $a_0/v_N \simeq 10^{-17} \ {\rm s} \ (v_N/10^{-2})^{-1}$ where $v_N$ is the nucleus final speed and $a_0$ is the Bohr radius. Such a perturbation is then non-adiabatic, $i.e.$ electrons do not respond to the sudden nuclear motion in a similar timescale, resulting in excitation or ionization \cite{Migdal:1977bq,Landau:1991wop}, the so-called Migdal effect which has been recently considered to extend the sensitivity of direct detection experiments \cite{Ibe:2017yqa,Dolan:2017xbu,Bernabei:2007jz,Moustakidis:2005gx,Ejiri:2005aj,Baxter:2019pnz,Knapen:2020aky}. In particular, numerical results from \cite{Ibe:2017yqa} indicate that the probability of outer-shell electron ionization for C and O atoms, the most abundant elements in IceCube and SNO+, is of order $f_e \simeq 10^{-2}-10^{-1}$ for the nuclear kinetic energies considered here, with the probability peak located at ionized electron energies $\sim 1-10 \ \rm eV$. Hence after this impulsive motion, a sizeable fraction of atoms are partially ionized. However, further considerations indicate the atoms will be fully ionized. The atoms accelerated to relative energies $100 \ {\rm eV} - 1 \ {\rm MeV}$, will scatter with the free electrons, resulting in further ionization. The cross-section for ionizing atomic oxygen or carbon is $\sigma_i \sim 10^{-16}-10^{-17} \ \rm cm^2$ in the energy range of interest \cite{1967ApJS...14..207L}. Ionization by electron-atom collisions will occur on a timescale given by $(f_e n_e v_N \sigma_i )^{-1} \lesssim 10^{-15} \ \rm s$, where $v_N$ is the atom velocity and $n_e \simeq 10^{23} \ \rm cm^{-3}$ is the electron number density. This timescale is shorter than the composite crossing time $(2R_X/v_X) \gtrsim 10^{-15} \ {\rm s} \ (R_X/{\rm nm}) (v_X/10^{-3})^{-1}$, so long as composites are larger than a nm, and becomes even shorter as more electrons are ionized and $f_e \sim 1$. Hence atoms are fully ionized in the detection regions shown in Fig.~\ref{fig:doxy}. 
These estimates agree with \cite{Massacrier:2011kt}, which finds order one ionization fractions for carbon and oxygen plasmas at $T \gtrsim 100 \ \rm eV$ and density $\sim 1 \ \rm g \ cm^{-3}$.

{\em Radiation.} The ion-electron plasma will have a photon opacity dominated by free-electron scattering, with a photon mean free path $(n_e \sigma_T)^{-1} \simeq 5 \ {\rm cm} \gg R_X$, where $\sigma_T \sim 10^{-24} \ \rm cm^{2}$ is the Thompson cross-section for electrons. Therefore, radiation never equilibrates with the plasma and we do not expect blackbody radiation. Instead, we expect thermal electron-ion bremsstrahlung, which has specific emissivity at frequency $\omega$ (see $e.g.$ \cite{padmanabhan_2000}) $j_{\omega}=(16\pi e^6 n_e^2/3 \sqrt{3} m_e^2) (2m_e/\pi T)^{1/2} \exp(-\omega/T)$, for electron mass $m_e$ and fine structure constant $e^2/4 \pi$. Since in Fig.~\ref{fig:doxy} we require $T \gtrsim 100 \ \rm eV$, there is emission of ionizing radiation. The integrated emissivity over volume and frequency yields a radiated energy rate 
\begin{align}
   \label{eq:ebrem}
   &\dot{E}_{\rm brem} = \frac{64\pi^2 e^6}{9 \sqrt{3} m_e^2} \left(\frac{2m_e T}{\pi}\right)^{\frac{1}{2}} n_e^2 R_X^3 \\
   &\simeq 10^{10} \ {\rm GeV \ s^{-1}}
   \left(\frac{g_X}{1}\right)^{-\frac{1}{2}} \left(\frac{g_n}{10^{-10}}\right)^{\frac{1}{2}} \left(\frac{m_X}{\rm TeV}\right)^{\frac{1}{2}} \left(\frac{R_X}{\rm nm}\right)^3. \nonumber
\end{align}
At temperatures $T \sim 100 \ {\rm keV-}1 \ \rm MeV$, we also expect a fraction of the ions to undergo thermonuclear fusion. In particular, we consider here the thermonuclear $^{16}$O burning rate tabulated in \cite{Caughlan:1987qf}, since this is the most abundant isotope in the terrestrial crust and mantle \cite{Dziewonski:1981xy,clarke1924composition,WANG2018460,Morgan6973,MCDONOUGH2003547,johnston1974}.
We remark that this radiation rate dominates over ionization energy losses.

{\em Detection.}  The large composites we have uncovered cannot be found by traditional dark matter experiments, which are flux limited to $M_X \lesssim 10^{19}$ GeV \cite{Bramante:2018qbc,Bramante:2018tos}. However, the copious energy released by fusion-capable composites make them observable at larger neutrino experiments like IceCube, Super-K, and large volume scintillators (LVS) like SNO+, Borexino, and JUNO; their enormity extends the $M_X$ mass reach to $3 \times 10^{25} \ \rm GeV$ in the case of IceCube (assuming 5 yrs and a ${\rm km^2}$ detection area). To conservatively establish the sensitivity of IceCube and LVS to a flood of $\gtrsim$ eV photons emitted from transiting composites, in Fig.~\ref{fig:doxy} we require trigger threshold energy depositions of $\sim 10 \ \rm TeV$ and $1 \ \rm MeV$ per $100 \ \rm ns$ respectively, 
which are an order of magnitude above the TeV \cite{Aartsen:2016nxy} and 100 keV \cite{Bellini:2013lnn,Andringa:2015tza} per 100 ns design thresholds of these experiments (this still underestimates IceCube's sensitivity, since our requirement implies $\gtrsim$ 100 PeV radiated in a transit through IceCube). Comparing this to Eq.~\eqref{eq:ebrem}, we find that nucleon couplings as small as $g_n \sim 10^{-14}$ at IceCube and $g_n \sim 10^{-12}$ at LVS can be detected in the upper left portions of IceCube and LVS detection regions marked in Fig.~\ref{fig:doxy}. Smaller coupling values will result in too little radiation rate for composite detection. We also show where $^{16}$O fusion reactions occur at IceCube as a single composite crosses it, using the $^{16}$O burning rate in \cite{Caughlan:1987qf}. In this case, there will be additional gamma rays and byproducts with $\sim$MeV energies, $e.g.$ $^{32,31}$S, $^{32}$P, $^{28}$Si, $^{24}$Mg as well as p, n and $\alpha$'s \cite{Caughlan:1987qf}.

{\em Capture on Earth and lack of X-nuclear scattering.} Thus far we have not mentioned nuclear scattering against $X$ fermions in composites. Compared to Eq.~\eqref{eq:ebrem}, composite energy loss from X-nuclear scattering will be negligible, in part because the fermi momentum of X is large, $p_{fX} \sim \bar m_X $. Accounting for nuclear scattering with degenerate fermions \cite{Joglekar:2019vzy,Joglekar:2020liw,Acevedo:2021kly}, the scattering energy loss is $\dot E_{X-N} \approx A^2 g_n^2 g_X^2 m_N^5 \bar{m}_{X}^{-4}(m_N+2\bar{m}_X) v_N^8$, which is tiny compared to bremsstrahlung in Fig.~\ref{fig:doxy}. On the other hand, energy loss in the form of radiation, $cf.$ Eq.~\eqref{eq:ebrem}, could result in stopping of composites before they reach detectors. This is relevant for lower mass composites with less initial kinetic energy. Using Eq.~\eqref{eq:ebrem}, a composite with an initial velocity $v_X$ will travel through the Earth's mantle a distance $L_{\rm cap}\simeq 2 \ {\rm km} \ (m_X/{\rm TeV})^{3/2}\,\, (10^{-10}/g_n)^{1/2}\,\, (1/g_X)^{3/2} \times \\  (v_X/200 \ {\rm km \ s^{-1}})^3\,\, (m_{\varphi}/{10\textrm{ keV}})^{2}$ before being slowed below Earth's escape velocity, where we have computed this distance considering the most abundant isotope $^{16}$O and using elemental/density profiles from \cite{Bramante:2019fhi,Acevedo:2020gro} (see also \cite{Dziewonski:1981xy,clarke1924composition,WANG2018460,Morgan6973,MCDONOUGH2003547,johnston1974,Mack:2007xj}). These scalings agree with the simple capture estimate $L_\text{cap}\sim\Delta E_\text{cap} v_\chi/\dot{E}_\text{brem}$, where $\Delta E_\text{cap}$ is the DM's initial kinetic energy. Earth's composite capture rate can be found using the method described in \cite{Acevedo:2020gro}. The captured composites may induce nuclear reactions in the crust and mantle, resulting in a potential planetary heat signal relevant for future searches \cite{Bramante:2019fhi,Acevedo:2020gro,Acevedo:2021kly}.

{\em WD explosions.} The transit of a large composite through a white dwarf (WD) can catalyze nuclear fusion reactions leading to a thermonuclear runaway and Type-Ia supernova explosion, similar to \cite{Bramante:2015cua,Graham:2015apa,Graham:2018efk,Acevedo:2019gre,Janish:2019nkk,Fedderke:2019jur}, although in this case fusion is initiated by nuclei accelerated inside the composite. As established in these references, WDs will ignite  when certain ignition conditions are met as detailed in \cite{1992Timmes}, where a set of critical temperatures and trigger masses are numerically computed for different white dwarf compositions and central densities. We conservatively require a critical temperature $T_{\rm crit}\simeq 1 \ {\rm MeV}$ for a pure $^{12}$C white dwarf. As they pass through a white dwarf, composites can lose kinetic energy to heat dissipation in the form of radiation, raising the possibility that they may be stopped before reaching the WD core. However, composites bounded by WDs in Fig.~\ref{fig:doxy} are so massive that a negligible fraction of their kinetic energy is lost to this dissipative effect. Heat conduction out of the composite is dominated by relativistic white dwarf electrons, with a rate $\dot{Q}_{\rm cond}\simeq 4 \pi^{2} T_{X}^{4}R_{X}/15 \kappa_{c} \rho_* \approx 10^{27} \ {\rm GeV \ s^{-1}} \ (\rho_{*}/10^{9} {\rm \ g \ cm^{-3}})^{4/15} (R_X/\rm{\mu m})$, where $\kappa_{c}\simeq {10^{-9} \ {\rm cm^{2} \ g^{-1}} \ (T_{*}/10^{7} K)^{2.8} \ (10^{9} ~{\rm g \ cm^{-3}}  /\rho_*)^{1.6}}$ is the conductive opacity of the relativistic white dwarf electrons \cite{Potekhin:1999yv}. Composite radiation, on the other hand, is $\dot{Q}_{\rm rad}= 4\pi R_X^2 \nabla (\sigma T^4)/\kappa_r \rho_* \simeq 16\pi R_X^2\sigma T^4 m_\varphi/\kappa_r \rho_* \simeq 10^{24} \ {\rm GeV \ s^{-1}} (R_X/{\rm \mu m})^2 (m_\varphi/{\rm keV})$, where $\kappa_r \simeq 10^{7} \ {\rm cm^2 \ g^{-1}} (T_*/10^7 \ \rm K)^{-7/2} (\rho_*/10^9 \ {\rm g \ cm^{-3}})$ is the white dwarf radiative opacity dominated by free-free electron transitions \cite{Kippenhahn:1994wva,Fedderke:2019jur}. We have assumed a blackbody energy density since the stellar material is highly opaque to photons.

The rate of carbon fusion in dense WD matter is $\dot{R}_{th}\simeq 10^{42} \ {\rm cm^{-3} \ s^{-1}} (\rho_*/10^9 \rm \ g \ cm^{-3})^2$ at $T_{\rm crit} \simeq 1 \ \rm MeV$, with an average energy release rate $Q\simeq +3$ MeV per reaction \cite{Gasques:2005ar}. This yields a nuclear energy release rate $\dot{Q}_{\rm fus}\simeq 4 \pi Q \dot{R}_{th} R_X^3 /3 \gtrsim 10^{28} \ {\rm GeV \ s^{-1}}(R_X/\rm{\mu m})^3$. Therefore, for composites with radii $R_X \gtrsim \rm \mu m$, the heat release from nuclear fusion greatly exceeds conductive and radiative losses, setting the conditions for a sustained thermonuclear runaway. We remark that stellar masses contained within radii $\gtrsim {\rm \mu m}$ are $\gtrsim 10^{-3} \ \rm g$, which are in agreement with the minimum trigger masses outlined in \cite{1992Timmes}. Fig.~\ref{fig:doxy} shows the region where $V_n\sim \rm MeV$ composites ignite a WD by simply passing through. Since one encounter would occur for composite masses $M_X \lesssim 10^{42} \ \rm GeV$ in a $\sim \rm Gyr$ timescale, the survival of $e.g.$ WD J160420.40 \cite{Acevedo:2019gre} implies constraints on nucleon couplings $g_n \lesssim 10^{-12} (m_X/10^8 \ \rm GeV)^{-1}$ in that region.

{\em Big Bang Nucleosynthesis.} It is natural to wonder whether BBN may constrain fusion-capable composites through over-production or disintegration of isotopes. An extensive analysis of fusion-capable DM composites on primordial abundances using relevant reaction rates ($e.g.$ \cite{Descouvemont:2004cw,Ando:2005cz,Jedamzik:2009uy}) will be the subject of future work. Here we remark that in the IceCube and LVS detection regions shown in Fig.~\ref{fig:doxy}, early universe composites seem unlikely to alter standard BBN abundance predictions. Constraints on $g_n$ imply that even for the maximum coupling allowed, composites will not change the temperature of the primordial plasma until redshift $z_X \lesssim 10^5 (A/1) (g_n/10^{-10}) (m_X/\rm TeV)$. However, by this redshift the baryon density will be significantly diluted according to $\Omega_b \rho_c (1+z_X)^3$. The average number of baryons inside composites will then be $4 \pi \Omega_b \rho_c (1+z_X)^3 R_X^3/3m_b \simeq 10^{-11} (m_X/{\rm TeV})^3 (g_n/10^{-10})^3 (R_X/{\rm nm})^3$ where $m_b$ is the baryon mass. Comparing this to Fig.~\ref{fig:doxy}, parameter space where large neutrino experiments have sensitivity, corresponds to composite sizes too small to have more than one baryon per composite, by the time a baryon inside a composite would be substantially accelerated in the early universe. Similar estimates using Eq.~\eqref{eq:ebrem} indicate that detectable fusion-capable composites do not observably alter the baryon-to-photon ratio after BBN, nor the ionization fraction after recombination.

\section{Conclusions}
We have studied the cosmology and detection of heavy composite DM that internally accelerates nuclei, resulting in copious collisional radiation and nuclear fusion. Prospects have been explored for detection of fusion-capable composites at IceCube and liquid scintillator experiments. There are many aspects of Standard Model particle acceleration in DM composites that remain. While here we considered composites that accelerate nuclei to MeV energies, if this were increased to relativistic energies, this would cause repulsive composite-SM scattering processes \cite{Greiner:1990tz}. For smaller than 100 eV acceleration energies, the Migdal effect and SM-SM collisional ionization should permit dark matter experiments to search for rather weakly-coupled composites. For liquid noble element experiments such as Xenon-1T, LUX, LZ or DEAP-3600, this will require a dedicated analysis of the scintillation signals produced and detection efficiencies \cite{Manzur:2009hp,Kimura:2019rdg,Dolan:2017xbu,McCabe:2017rln}. Given that asymmetric composites are often associated with SM asymmetries, similar acceleration effects should be explored for composites coupled to the SM through vector fields, and especially fields that couple to leptons, baryons, or a combination of these. Finally, it would be interesting to study whether fusion-capable composites could detectably alter isotopic abundances in the Earth over geological time periods. We leave these and other inquests into accelerative dark matter to future work.

\begin{acknowledgments}
\textit{Acknowledgements.} We thank Nirmal Raj for discussions and comments on the manuscript. The work of JA, JB, AG is supported by the Natural Sciences and Engineering Research Council of Canada (NSERC). Research at Perimeter Institute is supported in part by the Government of Canada through the Department of Innovation, Science and Economic Development Canada and by the Province of Ontario through the Ministry of Colleges and Universities.
\end{acknowledgments}

\bibliography{adfc}

\begin{thebibliography}{85}%
\makeatletter
\providecommand \@ifxundefined [1]{%
 \@ifx{#1\undefined}
}%
\providecommand \@ifnum [1]{%
 \ifnum #1\expandafter \@firstoftwo
 \else \expandafter \@secondoftwo
 \fi
}%
\providecommand \@ifx [1]{%
 \ifx #1\expandafter \@firstoftwo
 \else \expandafter \@secondoftwo
 \fi
}%
\providecommand \natexlab [1]{#1}%
\providecommand \enquote  [1]{``#1''}%
\providecommand \bibnamefont  [1]{#1}%
\providecommand \bibfnamefont [1]{#1}%
\providecommand \citenamefont [1]{#1}%
\providecommand \href@noop [0]{\@secondoftwo}%
\providecommand \href [0]{\begingroup \@sanitize@url \@href}%
\providecommand \@href[1]{\@@startlink{#1}\@@href}%
\providecommand \@@href[1]{\endgroup#1\@@endlink}%
\providecommand \@sanitize@url [0]{\catcode `\\12\catcode `\$12\catcode
  `\&12\catcode `\#12\catcode `\^12\catcode `\_12\catcode `\%12\relax}%
\providecommand \@@startlink[1]{}%
\providecommand \@@endlink[0]{}%
\providecommand \url  [0]{\begingroup\@sanitize@url \@url }%
\providecommand \@url [1]{\endgroup\@href {#1}{\urlprefix }}%
\providecommand \urlprefix  [0]{URL }%
\providecommand \Eprint [0]{\href }%
\providecommand \doibase [0]{https://doi.org/}%
\providecommand \selectlanguage [0]{\@gobble}%
\providecommand \bibinfo  [0]{\@secondoftwo}%
\providecommand \bibfield  [0]{\@secondoftwo}%
\providecommand \translation [1]{[#1]}%
\providecommand \BibitemOpen [0]{}%
\providecommand \bibitemStop [0]{}%
\providecommand \bibitemNoStop [0]{.\EOS\space}%
\providecommand \EOS [0]{\spacefactor3000\relax}%
\providecommand \BibitemShut  [1]{\csname bibitem#1\endcsname}%
\let\auto@bib@innerbib\@empty
\bibitem [{\citenamefont {Nussinov}(1985)}]{Nussinov:1985xr}%
  \BibitemOpen
  \bibfield  {author} {\bibinfo {author} {\bibfnamefont {S.}~\bibnamefont
  {Nussinov}},\ }\bibfield  {title} {\bibinfo {title} {{Technocosmology: could
  a technibaryon excess provide a 'natural' missing mass candidate?}},\ }\href
  {https://doi.org/10.1016/0370-2693(85)90689-6} {\bibfield  {journal}
  {\bibinfo  {journal} {Phys. Lett. B}\ }\textbf {\bibinfo {volume} {165}},\
  \bibinfo {pages} {55} (\bibinfo {year} {1985})}\BibitemShut {NoStop}%
\bibitem [{\citenamefont {Bagnasco}\ \emph {et~al.}(1994)\citenamefont
  {Bagnasco}, \citenamefont {Dine},\ and\ \citenamefont
  {Thomas}}]{Bagnasco:1993st}%
  \BibitemOpen
  \bibfield  {author} {\bibinfo {author} {\bibfnamefont {J.}~\bibnamefont
  {Bagnasco}}, \bibinfo {author} {\bibfnamefont {M.}~\bibnamefont {Dine}},\
  and\ \bibinfo {author} {\bibfnamefont {S.~D.}\ \bibnamefont {Thomas}},\
  }\bibfield  {title} {\bibinfo {title} {{Detecting technibaryon dark
  matter}},\ }\href {https://doi.org/10.1016/0370-2693(94)90830-3} {\bibfield
  {journal} {\bibinfo  {journal} {Phys. Lett. B}\ }\textbf {\bibinfo {volume}
  {320}},\ \bibinfo {pages} {99} (\bibinfo {year} {1994})},\ \Eprint
  {https://arxiv.org/abs/hep-ph/9310290} {arXiv:hep-ph/9310290} \BibitemShut
  {NoStop}%
\bibitem [{\citenamefont {Alves}\ \emph {et~al.}(2010)\citenamefont {Alves},
  \citenamefont {Behbahani}, \citenamefont {Schuster},\ and\ \citenamefont
  {Wacker}}]{Alves:2009nf}%
  \BibitemOpen
  \bibfield  {author} {\bibinfo {author} {\bibfnamefont {D.~S.~M.}\
  \bibnamefont {Alves}}, \bibinfo {author} {\bibfnamefont {S.~R.}\ \bibnamefont
  {Behbahani}}, \bibinfo {author} {\bibfnamefont {P.}~\bibnamefont
  {Schuster}},\ and\ \bibinfo {author} {\bibfnamefont {J.~G.}\ \bibnamefont
  {Wacker}},\ }\bibfield  {title} {\bibinfo {title} {{Composite Inelastic Dark
  Matter}},\ }\href {https://doi.org/10.1016/j.physletb.2010.08.006} {\bibfield
   {journal} {\bibinfo  {journal} {Phys. Lett.}\ }\textbf {\bibinfo {volume}
  {B692}},\ \bibinfo {pages} {323} (\bibinfo {year} {2010})},\ \Eprint
  {https://arxiv.org/abs/0903.3945} {arXiv:0903.3945 [hep-ph]} \BibitemShut
  {NoStop}%
\bibitem [{\citenamefont {Kribs}\ \emph {et~al.}(2010)\citenamefont {Kribs},
  \citenamefont {Roy}, \citenamefont {Terning},\ and\ \citenamefont
  {Zurek}}]{Kribs:2009fy}%
  \BibitemOpen
  \bibfield  {author} {\bibinfo {author} {\bibfnamefont {G.~D.}\ \bibnamefont
  {Kribs}}, \bibinfo {author} {\bibfnamefont {T.~S.}\ \bibnamefont {Roy}},
  \bibinfo {author} {\bibfnamefont {J.}~\bibnamefont {Terning}},\ and\ \bibinfo
  {author} {\bibfnamefont {K.~M.}\ \bibnamefont {Zurek}},\ }\bibfield  {title}
  {\bibinfo {title} {{Quirky Composite Dark Matter}},\ }\href
  {https://doi.org/10.1103/PhysRevD.81.095001} {\bibfield  {journal} {\bibinfo
  {journal} {Phys. Rev. D}\ }\textbf {\bibinfo {volume} {81}},\ \bibinfo
  {pages} {095001} (\bibinfo {year} {2010})},\ \Eprint
  {https://arxiv.org/abs/0909.2034} {arXiv:0909.2034 [hep-ph]} \BibitemShut
  {NoStop}%
\bibitem [{\citenamefont {Lee}\ \emph {et~al.}(2014)\citenamefont {Lee},
  \citenamefont {Park},\ and\ \citenamefont {Sanz}}]{Lee:2013bua}%
  \BibitemOpen
  \bibfield  {author} {\bibinfo {author} {\bibfnamefont {H.~M.}\ \bibnamefont
  {Lee}}, \bibinfo {author} {\bibfnamefont {M.}~\bibnamefont {Park}},\ and\
  \bibinfo {author} {\bibfnamefont {V.}~\bibnamefont {Sanz}},\ }\bibfield
  {title} {\bibinfo {title} {{Gravity-mediated (or Composite) Dark Matter}},\
  }\href {https://doi.org/10.1140/epjc/s10052-014-2715-8} {\bibfield  {journal}
  {\bibinfo  {journal} {Eur. Phys. J. C}\ }\textbf {\bibinfo {volume} {74}},\
  \bibinfo {pages} {2715} (\bibinfo {year} {2014})},\ \Eprint
  {https://arxiv.org/abs/1306.4107} {arXiv:1306.4107 [hep-ph]} \BibitemShut
  {NoStop}%
\bibitem [{\citenamefont {Krnjaic}\ and\ \citenamefont
  {Sigurdson}(2015)}]{Krnjaic:2014xza}%
  \BibitemOpen
  \bibfield  {author} {\bibinfo {author} {\bibfnamefont {G.}~\bibnamefont
  {Krnjaic}}\ and\ \bibinfo {author} {\bibfnamefont {K.}~\bibnamefont
  {Sigurdson}},\ }\bibfield  {title} {\bibinfo {title} {{Big Bang
  Darkleosynthesis}},\ }\href {https://doi.org/10.1016/j.physletb.2015.11.001}
  {\bibfield  {journal} {\bibinfo  {journal} {Phys. Lett.}\ }\textbf {\bibinfo
  {volume} {B751}},\ \bibinfo {pages} {464} (\bibinfo {year} {2015})},\ \Eprint
  {https://arxiv.org/abs/1406.1171} {arXiv:1406.1171 [hep-ph]} \BibitemShut
  {NoStop}%
\bibitem [{\citenamefont {Detmold}\ \emph {et~al.}(2014)\citenamefont
  {Detmold}, \citenamefont {McCullough},\ and\ \citenamefont
  {Pochinsky}}]{Detmold:2014qqa}%
  \BibitemOpen
  \bibfield  {author} {\bibinfo {author} {\bibfnamefont {W.}~\bibnamefont
  {Detmold}}, \bibinfo {author} {\bibfnamefont {M.}~\bibnamefont
  {McCullough}},\ and\ \bibinfo {author} {\bibfnamefont {A.}~\bibnamefont
  {Pochinsky}},\ }\bibfield  {title} {\bibinfo {title} {{Dark Nuclei I:
  Cosmology and Indirect Detection}},\ }\href
  {https://doi.org/10.1103/PhysRevD.90.115013} {\bibfield  {journal} {\bibinfo
  {journal} {Phys. Rev.}\ }\textbf {\bibinfo {volume} {D90}},\ \bibinfo {pages}
  {115013} (\bibinfo {year} {2014})},\ \Eprint
  {https://arxiv.org/abs/1406.2276} {arXiv:1406.2276 [hep-ph]} \BibitemShut
  {NoStop}%
\bibitem [{\citenamefont {Jacobs}\ \emph {et~al.}(2015)\citenamefont {Jacobs},
  \citenamefont {Starkman},\ and\ \citenamefont {Lynn}}]{Jacobs:2014yca}%
  \BibitemOpen
  \bibfield  {author} {\bibinfo {author} {\bibfnamefont {D.~M.}\ \bibnamefont
  {Jacobs}}, \bibinfo {author} {\bibfnamefont {G.~D.}\ \bibnamefont
  {Starkman}},\ and\ \bibinfo {author} {\bibfnamefont {B.~W.}\ \bibnamefont
  {Lynn}},\ }\bibfield  {title} {\bibinfo {title} {{Macro Dark Matter}},\
  }\href {https://doi.org/10.1093/mnras/stv774} {\bibfield  {journal} {\bibinfo
   {journal} {Mon. Not. Roy. Astron. Soc.}\ }\textbf {\bibinfo {volume}
  {450}},\ \bibinfo {pages} {3418} (\bibinfo {year} {2015})},\ \Eprint
  {https://arxiv.org/abs/1410.2236} {arXiv:1410.2236 [astro-ph.CO]}
  \BibitemShut {NoStop}%
\bibitem [{\citenamefont {Bramante}\ \emph
  {et~al.}(2019{\natexlab{a}})\citenamefont {Bramante}, \citenamefont
  {Broerman}, \citenamefont {Kumar}, \citenamefont {Lang}, \citenamefont
  {Pospelov},\ and\ \citenamefont {Raj}}]{Bramante:2018tos}%
  \BibitemOpen
  \bibfield  {author} {\bibinfo {author} {\bibfnamefont {J.}~\bibnamefont
  {Bramante}}, \bibinfo {author} {\bibfnamefont {B.}~\bibnamefont {Broerman}},
  \bibinfo {author} {\bibfnamefont {J.}~\bibnamefont {Kumar}}, \bibinfo
  {author} {\bibfnamefont {R.~F.}\ \bibnamefont {Lang}}, \bibinfo {author}
  {\bibfnamefont {M.}~\bibnamefont {Pospelov}},\ and\ \bibinfo {author}
  {\bibfnamefont {N.}~\bibnamefont {Raj}},\ }\bibfield  {title} {\bibinfo
  {title} {{Foraging for dark matter in large volume liquid scintillator
  neutrino detectors with multiscatter events}},\ }\href
  {https://doi.org/10.1103/PhysRevD.99.083010} {\bibfield  {journal} {\bibinfo
  {journal} {Phys. Rev.}\ }\textbf {\bibinfo {volume} {D99}},\ \bibinfo {pages}
  {083010} (\bibinfo {year} {2019}{\natexlab{a}})},\ \Eprint
  {https://arxiv.org/abs/1812.09325} {arXiv:1812.09325 [hep-ph]} \BibitemShut
  {NoStop}%
\bibitem [{\citenamefont {Ibe}\ \emph {et~al.}(2018{\natexlab{a}})\citenamefont
  {Ibe}, \citenamefont {Kamada}, \citenamefont {Kobayashi},\ and\ \citenamefont
  {Nakano}}]{Ibe:2018juk}%
  \BibitemOpen
  \bibfield  {author} {\bibinfo {author} {\bibfnamefont {M.}~\bibnamefont
  {Ibe}}, \bibinfo {author} {\bibfnamefont {A.}~\bibnamefont {Kamada}},
  \bibinfo {author} {\bibfnamefont {S.}~\bibnamefont {Kobayashi}},\ and\
  \bibinfo {author} {\bibfnamefont {W.}~\bibnamefont {Nakano}},\ }\bibfield
  {title} {\bibinfo {title} {{Composite Asymmetric Dark Matter with a Dark
  Photon Portal}},\ }\href {https://doi.org/10.1007/JHEP11(2018)203} {\bibfield
   {journal} {\bibinfo  {journal} {JHEP}\ }\textbf {\bibinfo {volume} {11}},\
  \bibinfo {pages} {203}},\ \Eprint {https://arxiv.org/abs/1805.06876}
  {arXiv:1805.06876 [hep-ph]} \BibitemShut {NoStop}%
\bibitem [{\citenamefont {Coskuner}\ \emph {et~al.}(2019)\citenamefont
  {Coskuner}, \citenamefont {Grabowska}, \citenamefont {Knapen},\ and\
  \citenamefont {Zurek}}]{Coskuner:2018are}%
  \BibitemOpen
  \bibfield  {author} {\bibinfo {author} {\bibfnamefont {A.}~\bibnamefont
  {Coskuner}}, \bibinfo {author} {\bibfnamefont {D.~M.}\ \bibnamefont
  {Grabowska}}, \bibinfo {author} {\bibfnamefont {S.}~\bibnamefont {Knapen}},\
  and\ \bibinfo {author} {\bibfnamefont {K.~M.}\ \bibnamefont {Zurek}},\
  }\bibfield  {title} {\bibinfo {title} {{Direct Detection of Bound States of
  Asymmetric Dark Matter}},\ }\href
  {https://doi.org/10.1103/PhysRevD.100.035025} {\bibfield  {journal} {\bibinfo
   {journal} {Phys. Rev.}\ }\textbf {\bibinfo {volume} {D100}},\ \bibinfo
  {pages} {035025} (\bibinfo {year} {2019})},\ \Eprint
  {https://arxiv.org/abs/1812.07573} {arXiv:1812.07573 [hep-ph]} \BibitemShut
  {NoStop}%
\bibitem [{\citenamefont {Bai}\ \emph {et~al.}(2019)\citenamefont {Bai},
  \citenamefont {Long},\ and\ \citenamefont {Lu}}]{Bai:2018dxf}%
  \BibitemOpen
  \bibfield  {author} {\bibinfo {author} {\bibfnamefont {Y.}~\bibnamefont
  {Bai}}, \bibinfo {author} {\bibfnamefont {A.~J.}\ \bibnamefont {Long}},\ and\
  \bibinfo {author} {\bibfnamefont {S.}~\bibnamefont {Lu}},\ }\bibfield
  {title} {\bibinfo {title} {{Dark Quark Nuggets}},\ }\href
  {https://doi.org/10.1103/PhysRevD.99.055047} {\bibfield  {journal} {\bibinfo
  {journal} {Phys. Rev. D}\ }\textbf {\bibinfo {volume} {99}},\ \bibinfo
  {pages} {055047} (\bibinfo {year} {2019})},\ \Eprint
  {https://arxiv.org/abs/1810.04360} {arXiv:1810.04360 [hep-ph]} \BibitemShut
  {NoStop}%
\bibitem [{\citenamefont {Bai}\ and\ \citenamefont
  {Berger}(2020)}]{Bai:2019ogh}%
  \BibitemOpen
  \bibfield  {author} {\bibinfo {author} {\bibfnamefont {Y.}~\bibnamefont
  {Bai}}\ and\ \bibinfo {author} {\bibfnamefont {J.}~\bibnamefont {Berger}},\
  }\bibfield  {title} {\bibinfo {title} {{Nucleus Capture by Macroscopic Dark
  Matter}},\ }\href {https://doi.org/10.1007/JHEP05(2020)160} {\bibfield
  {journal} {\bibinfo  {journal} {JHEP}\ }\textbf {\bibinfo {volume} {05}},\
  \bibinfo {pages} {160}},\ \Eprint {https://arxiv.org/abs/1912.02813}
  {arXiv:1912.02813 [hep-ph]} \BibitemShut {NoStop}%
\bibitem [{\citenamefont {Bramante}\ \emph
  {et~al.}(2019{\natexlab{b}})\citenamefont {Bramante}, \citenamefont {Kumar},\
  and\ \citenamefont {Raj}}]{Bramante:2019yss}%
  \BibitemOpen
  \bibfield  {author} {\bibinfo {author} {\bibfnamefont {J.}~\bibnamefont
  {Bramante}}, \bibinfo {author} {\bibfnamefont {J.}~\bibnamefont {Kumar}},\
  and\ \bibinfo {author} {\bibfnamefont {N.}~\bibnamefont {Raj}},\ }\bibfield
  {title} {\bibinfo {title} {{Dark matter astrometry at underground detectors
  with multiscatter events}},\ }\href
  {https://doi.org/10.1103/PhysRevD.100.123016} {\bibfield  {journal} {\bibinfo
   {journal} {Phys. Rev. D}\ }\textbf {\bibinfo {volume} {100}},\ \bibinfo
  {pages} {123016} (\bibinfo {year} {2019}{\natexlab{b}})},\ \Eprint
  {https://arxiv.org/abs/1910.05380} {arXiv:1910.05380 [hep-ph]} \BibitemShut
  {NoStop}%
\bibitem [{\citenamefont {Wise}\ and\ \citenamefont
  {Zhang}(2014)}]{Wise:2014jva}%
  \BibitemOpen
  \bibfield  {author} {\bibinfo {author} {\bibfnamefont {M.~B.}\ \bibnamefont
  {Wise}}\ and\ \bibinfo {author} {\bibfnamefont {Y.}~\bibnamefont {Zhang}},\
  }\bibfield  {title} {\bibinfo {title} {{Stable Bound States of Asymmetric
  Dark Matter}},\ }\href {https://doi.org/10.1103/PhysRevD.90.055030}
  {\bibfield  {journal} {\bibinfo  {journal} {Phys. Rev. D}\ }\textbf {\bibinfo
  {volume} {90}},\ \bibinfo {pages} {055030} (\bibinfo {year} {2014})},\
  \bibinfo {note} {[Erratum: Phys.Rev.D 91, 039907 (2015)]},\ \Eprint
  {https://arxiv.org/abs/1407.4121} {arXiv:1407.4121 [hep-ph]} \BibitemShut
  {NoStop}%
\bibitem [{\citenamefont {Wise}\ and\ \citenamefont
  {Zhang}(2015)}]{Wise:2014ola}%
  \BibitemOpen
  \bibfield  {author} {\bibinfo {author} {\bibfnamefont {M.~B.}\ \bibnamefont
  {Wise}}\ and\ \bibinfo {author} {\bibfnamefont {Y.}~\bibnamefont {Zhang}},\
  }\bibfield  {title} {\bibinfo {title} {{Yukawa Bound States of a Large Number
  of Fermions}},\ }\href {https://doi.org/10.1007/JHEP10(2015)165,
  10.1007/JHEP02(2015)023} {\bibfield  {journal} {\bibinfo  {journal} {JHEP}\
  }\textbf {\bibinfo {volume} {02}},\ \bibinfo {pages} {023}},\ \bibinfo {note}
  {[Erratum: JHEP10,165(2015)]},\ \Eprint {https://arxiv.org/abs/1411.1772}
  {arXiv:1411.1772 [hep-ph]} \BibitemShut {NoStop}%
\bibitem [{\citenamefont {Hardy}\ \emph
  {et~al.}(2015{\natexlab{a}})\citenamefont {Hardy}, \citenamefont {Lasenby},
  \citenamefont {March-Russell},\ and\ \citenamefont {West}}]{Hardy:2014mqa}%
  \BibitemOpen
  \bibfield  {author} {\bibinfo {author} {\bibfnamefont {E.}~\bibnamefont
  {Hardy}}, \bibinfo {author} {\bibfnamefont {R.}~\bibnamefont {Lasenby}},
  \bibinfo {author} {\bibfnamefont {J.}~\bibnamefont {March-Russell}},\ and\
  \bibinfo {author} {\bibfnamefont {S.~M.}\ \bibnamefont {West}},\ }\bibfield
  {title} {\bibinfo {title} {{Big Bang Synthesis of Nuclear Dark Matter}},\
  }\href {https://doi.org/10.1007/JHEP06(2015)011} {\bibfield  {journal}
  {\bibinfo  {journal} {JHEP}\ }\textbf {\bibinfo {volume} {06}},\ \bibinfo
  {pages} {011}},\ \Eprint {https://arxiv.org/abs/1411.3739} {arXiv:1411.3739
  [hep-ph]} \BibitemShut {NoStop}%
\bibitem [{\citenamefont {Hardy}\ \emph
  {et~al.}(2015{\natexlab{b}})\citenamefont {Hardy}, \citenamefont {Lasenby},
  \citenamefont {March-Russell},\ and\ \citenamefont {West}}]{Hardy:2015boa}%
  \BibitemOpen
  \bibfield  {author} {\bibinfo {author} {\bibfnamefont {E.}~\bibnamefont
  {Hardy}}, \bibinfo {author} {\bibfnamefont {R.}~\bibnamefont {Lasenby}},
  \bibinfo {author} {\bibfnamefont {J.}~\bibnamefont {March-Russell}},\ and\
  \bibinfo {author} {\bibfnamefont {S.~M.}\ \bibnamefont {West}},\ }\bibfield
  {title} {\bibinfo {title} {{Signatures of Large Composite Dark Matter
  States}},\ }\href {https://doi.org/10.1007/JHEP07(2015)133} {\bibfield
  {journal} {\bibinfo  {journal} {JHEP}\ }\textbf {\bibinfo {volume} {07}},\
  \bibinfo {pages} {133}},\ \Eprint {https://arxiv.org/abs/1504.05419}
  {arXiv:1504.05419 [hep-ph]} \BibitemShut {NoStop}%
\bibitem [{\citenamefont {Gresham}\ \emph {et~al.}(2017)\citenamefont
  {Gresham}, \citenamefont {Lou},\ and\ \citenamefont
  {Zurek}}]{Gresham:2017zqi}%
  \BibitemOpen
  \bibfield  {author} {\bibinfo {author} {\bibfnamefont {M.~I.}\ \bibnamefont
  {Gresham}}, \bibinfo {author} {\bibfnamefont {H.~K.}\ \bibnamefont {Lou}},\
  and\ \bibinfo {author} {\bibfnamefont {K.~M.}\ \bibnamefont {Zurek}},\
  }\bibfield  {title} {\bibinfo {title} {{Nuclear Structure of Bound States of
  Asymmetric Dark Matter}},\ }\href
  {https://doi.org/10.1103/PhysRevD.96.096012} {\bibfield  {journal} {\bibinfo
  {journal} {Phys. Rev.}\ }\textbf {\bibinfo {volume} {D96}},\ \bibinfo {pages}
  {096012} (\bibinfo {year} {2017})},\ \Eprint
  {https://arxiv.org/abs/1707.02313} {arXiv:1707.02313 [hep-ph]} \BibitemShut
  {NoStop}%
\bibitem [{\citenamefont {Gresham}\ \emph
  {et~al.}(2018{\natexlab{a}})\citenamefont {Gresham}, \citenamefont {Lou},\
  and\ \citenamefont {Zurek}}]{Gresham:2017cvl}%
  \BibitemOpen
  \bibfield  {author} {\bibinfo {author} {\bibfnamefont {M.~I.}\ \bibnamefont
  {Gresham}}, \bibinfo {author} {\bibfnamefont {H.~K.}\ \bibnamefont {Lou}},\
  and\ \bibinfo {author} {\bibfnamefont {K.~M.}\ \bibnamefont {Zurek}},\
  }\bibfield  {title} {\bibinfo {title} {{Early Universe synthesis of
  asymmetric dark matter nuggets}},\ }\href
  {https://doi.org/10.1103/PhysRevD.97.036003} {\bibfield  {journal} {\bibinfo
  {journal} {Phys. Rev.}\ }\textbf {\bibinfo {volume} {D97}},\ \bibinfo {pages}
  {036003} (\bibinfo {year} {2018}{\natexlab{a}})},\ \Eprint
  {https://arxiv.org/abs/1707.02316} {arXiv:1707.02316 [hep-ph]} \BibitemShut
  {NoStop}%
\bibitem [{\citenamefont {Gresham}\ \emph
  {et~al.}(2018{\natexlab{b}})\citenamefont {Gresham}, \citenamefont {Lou},\
  and\ \citenamefont {Zurek}}]{Gresham:2018anj}%
  \BibitemOpen
  \bibfield  {author} {\bibinfo {author} {\bibfnamefont {M.~I.}\ \bibnamefont
  {Gresham}}, \bibinfo {author} {\bibfnamefont {H.~K.}\ \bibnamefont {Lou}},\
  and\ \bibinfo {author} {\bibfnamefont {K.~M.}\ \bibnamefont {Zurek}},\
  }\bibfield  {title} {\bibinfo {title} {{Astrophysical Signatures of
  Asymmetric Dark Matter Bound States}},\ }\href
  {https://doi.org/10.1103/PhysRevD.98.096001} {\bibfield  {journal} {\bibinfo
  {journal} {Phys. Rev.}\ }\textbf {\bibinfo {volume} {D98}},\ \bibinfo {pages}
  {096001} (\bibinfo {year} {2018}{\natexlab{b}})},\ \Eprint
  {https://arxiv.org/abs/1805.04512} {arXiv:1805.04512 [hep-ph]} \BibitemShut
  {NoStop}%
\bibitem [{\citenamefont {Affleck}\ and\ \citenamefont
  {Dine}(1985)}]{Affleck:1984fy}%
  \BibitemOpen
  \bibfield  {author} {\bibinfo {author} {\bibfnamefont {I.}~\bibnamefont
  {Affleck}}\ and\ \bibinfo {author} {\bibfnamefont {M.}~\bibnamefont {Dine}},\
  }\bibfield  {title} {\bibinfo {title} {{A New Mechanism for Baryogenesis}},\
  }\href {https://doi.org/10.1016/0550-3213(85)90021-5} {\bibfield  {journal}
  {\bibinfo  {journal} {Nucl. Phys.}\ }\textbf {\bibinfo {volume} {B249}},\
  \bibinfo {pages} {361} (\bibinfo {year} {1985})}\BibitemShut {NoStop}%
\bibitem [{\citenamefont {Dine}\ \emph {et~al.}(1996)\citenamefont {Dine},
  \citenamefont {Randall},\ and\ \citenamefont {Thomas}}]{Dine:1995kz}%
  \BibitemOpen
  \bibfield  {author} {\bibinfo {author} {\bibfnamefont {M.}~\bibnamefont
  {Dine}}, \bibinfo {author} {\bibfnamefont {L.}~\bibnamefont {Randall}},\ and\
  \bibinfo {author} {\bibfnamefont {S.~D.}\ \bibnamefont {Thomas}},\ }\bibfield
   {title} {\bibinfo {title} {{Baryogenesis from flat directions of the
  supersymmetric standard model}},\ }\href
  {https://doi.org/10.1016/0550-3213(95)00538-2} {\bibfield  {journal}
  {\bibinfo  {journal} {Nucl. Phys.}\ }\textbf {\bibinfo {volume} {B458}},\
  \bibinfo {pages} {291} (\bibinfo {year} {1996})},\ \Eprint
  {https://arxiv.org/abs/hep-ph/9507453} {arXiv:hep-ph/9507453 [hep-ph]}
  \BibitemShut {NoStop}%
\bibitem [{\citenamefont {Bramante}\ and\ \citenamefont
  {Unwin}(2017)}]{Bramante:2017obj}%
  \BibitemOpen
  \bibfield  {author} {\bibinfo {author} {\bibfnamefont {J.}~\bibnamefont
  {Bramante}}\ and\ \bibinfo {author} {\bibfnamefont {J.}~\bibnamefont
  {Unwin}},\ }\bibfield  {title} {\bibinfo {title} {{Superheavy Thermal Dark
  Matter and Primordial Asymmetries}},\ }\href
  {https://doi.org/10.1007/JHEP02(2017)119} {\bibfield  {journal} {\bibinfo
  {journal} {JHEP}\ }\textbf {\bibinfo {volume} {02}},\ \bibinfo {pages}
  {119}},\ \Eprint {https://arxiv.org/abs/1701.05859} {arXiv:1701.05859
  [hep-ph]} \BibitemShut {NoStop}%
\bibitem [{\citenamefont {Zurek}(2014)}]{Zurek:2013wia}%
  \BibitemOpen
  \bibfield  {author} {\bibinfo {author} {\bibfnamefont {K.~M.}\ \bibnamefont
  {Zurek}},\ }\bibfield  {title} {\bibinfo {title} {{Asymmetric Dark Matter:
  Theories, Signatures, and Constraints}},\ }\href
  {https://doi.org/10.1016/j.physrep.2013.12.001} {\bibfield  {journal}
  {\bibinfo  {journal} {Phys. Rept.}\ }\textbf {\bibinfo {volume} {537}},\
  \bibinfo {pages} {91} (\bibinfo {year} {2014})},\ \Eprint
  {https://arxiv.org/abs/1308.0338} {arXiv:1308.0338 [hep-ph]} \BibitemShut
  {NoStop}%
\bibitem [{\citenamefont {Petraki}\ and\ \citenamefont
  {Volkas}(2013)}]{Petraki:2013wwa}%
  \BibitemOpen
  \bibfield  {author} {\bibinfo {author} {\bibfnamefont {K.}~\bibnamefont
  {Petraki}}\ and\ \bibinfo {author} {\bibfnamefont {R.~R.}\ \bibnamefont
  {Volkas}},\ }\bibfield  {title} {\bibinfo {title} {{Review of asymmetric dark
  matter}},\ }\href {https://doi.org/10.1142/S0217751X13300287} {\bibfield
  {journal} {\bibinfo  {journal} {Int. J. Mod. Phys.}\ }\textbf {\bibinfo
  {volume} {A28}},\ \bibinfo {pages} {1330028} (\bibinfo {year} {2013})},\
  \Eprint {https://arxiv.org/abs/1305.4939} {arXiv:1305.4939 [hep-ph]}
  \BibitemShut {NoStop}%
\bibitem [{\citenamefont {Banks}\ \emph {et~al.}(1994)\citenamefont {Banks},
  \citenamefont {Kaplan},\ and\ \citenamefont {Nelson}}]{Banks:1993en}%
  \BibitemOpen
  \bibfield  {author} {\bibinfo {author} {\bibfnamefont {T.}~\bibnamefont
  {Banks}}, \bibinfo {author} {\bibfnamefont {D.~B.}\ \bibnamefont {Kaplan}},\
  and\ \bibinfo {author} {\bibfnamefont {A.~E.}\ \bibnamefont {Nelson}},\
  }\bibfield  {title} {\bibinfo {title} {{Cosmological implications of
  dynamical supersymmetry breaking}},\ }\href
  {https://doi.org/10.1103/PhysRevD.49.779} {\bibfield  {journal} {\bibinfo
  {journal} {Phys. Rev. D}\ }\textbf {\bibinfo {volume} {49}},\ \bibinfo
  {pages} {779} (\bibinfo {year} {1994})},\ \Eprint
  {https://arxiv.org/abs/hep-ph/9308292} {arXiv:hep-ph/9308292} \BibitemShut
  {NoStop}%
\bibitem [{\citenamefont {Randall}\ \emph {et~al.}(2016)\citenamefont
  {Randall}, \citenamefont {Scholtz},\ and\ \citenamefont
  {Unwin}}]{Randall:2015xza}%
  \BibitemOpen
  \bibfield  {author} {\bibinfo {author} {\bibfnamefont {L.}~\bibnamefont
  {Randall}}, \bibinfo {author} {\bibfnamefont {J.}~\bibnamefont {Scholtz}},\
  and\ \bibinfo {author} {\bibfnamefont {J.}~\bibnamefont {Unwin}},\ }\bibfield
   {title} {\bibinfo {title} {{Flooded Dark Matter and S Level Rise}},\ }\href
  {https://doi.org/10.1007/JHEP03(2016)011} {\bibfield  {journal} {\bibinfo
  {journal} {JHEP}\ }\textbf {\bibinfo {volume} {03}},\ \bibinfo {pages}
  {011}},\ \Eprint {https://arxiv.org/abs/1509.08477} {arXiv:1509.08477
  [hep-ph]} \BibitemShut {NoStop}%
\bibitem [{\citenamefont {Berlin}\ \emph {et~al.}(2016)\citenamefont {Berlin},
  \citenamefont {Hooper},\ and\ \citenamefont {Krnjaic}}]{Berlin:2016vnh}%
  \BibitemOpen
  \bibfield  {author} {\bibinfo {author} {\bibfnamefont {A.}~\bibnamefont
  {Berlin}}, \bibinfo {author} {\bibfnamefont {D.}~\bibnamefont {Hooper}},\
  and\ \bibinfo {author} {\bibfnamefont {G.}~\bibnamefont {Krnjaic}},\
  }\bibfield  {title} {\bibinfo {title} {{PeV-Scale Dark Matter as a Thermal
  Relic of a Decoupled Sector}},\ }\href
  {https://doi.org/10.1016/j.physletb.2016.06.037} {\bibfield  {journal}
  {\bibinfo  {journal} {Phys. Lett. B}\ }\textbf {\bibinfo {volume} {760}},\
  \bibinfo {pages} {106} (\bibinfo {year} {2016})},\ \Eprint
  {https://arxiv.org/abs/1602.08490} {arXiv:1602.08490 [hep-ph]} \BibitemShut
  {NoStop}%
\bibitem [{\citenamefont {Bernal}\ \emph {et~al.}(2019)\citenamefont {Bernal},
  \citenamefont {Elahi}, \citenamefont {Maldonado},\ and\ \citenamefont
  {Unwin}}]{Bernal:2019mhf}%
  \BibitemOpen
  \bibfield  {author} {\bibinfo {author} {\bibfnamefont {N.}~\bibnamefont
  {Bernal}}, \bibinfo {author} {\bibfnamefont {F.}~\bibnamefont {Elahi}},
  \bibinfo {author} {\bibfnamefont {C.}~\bibnamefont {Maldonado}},\ and\
  \bibinfo {author} {\bibfnamefont {J.}~\bibnamefont {Unwin}},\ }\bibfield
  {title} {\bibinfo {title} {{Ultraviolet Freeze-in and Non-Standard
  Cosmologies}},\ }\href {https://doi.org/10.1088/1475-7516/2019/11/026}
  {\bibfield  {journal} {\bibinfo  {journal} {JCAP}\ }\textbf {\bibinfo
  {volume} {11}},\ \bibinfo {pages} {026}},\ \Eprint
  {https://arxiv.org/abs/1909.07992} {arXiv:1909.07992 [hep-ph]} \BibitemShut
  {NoStop}%
\bibitem [{\citenamefont {Evans}\ \emph {et~al.}(2020)\citenamefont {Evans},
  \citenamefont {Ghalsasi}, \citenamefont {Gori}, \citenamefont {Tammaro},\
  and\ \citenamefont {Zupan}}]{Evans:2019jcs}%
  \BibitemOpen
  \bibfield  {author} {\bibinfo {author} {\bibfnamefont {J.~A.}\ \bibnamefont
  {Evans}}, \bibinfo {author} {\bibfnamefont {A.}~\bibnamefont {Ghalsasi}},
  \bibinfo {author} {\bibfnamefont {S.}~\bibnamefont {Gori}}, \bibinfo {author}
  {\bibfnamefont {M.}~\bibnamefont {Tammaro}},\ and\ \bibinfo {author}
  {\bibfnamefont {J.}~\bibnamefont {Zupan}},\ }\bibfield  {title} {\bibinfo
  {title} {{Light Dark Matter from Entropy Dilution}},\ }\href
  {https://doi.org/10.1007/JHEP02(2020)151} {\bibfield  {journal} {\bibinfo
  {journal} {JHEP}\ }\textbf {\bibinfo {volume} {02}},\ \bibinfo {pages}
  {151}},\ \Eprint {https://arxiv.org/abs/1910.06319} {arXiv:1910.06319
  [hep-ph]} \BibitemShut {NoStop}%
\bibitem [{\citenamefont {Burgess}\ \emph {et~al.}(2005)\citenamefont
  {Burgess}, \citenamefont {Easther}, \citenamefont {Mazumdar}, \citenamefont
  {Mota},\ and\ \citenamefont {Multamaki}}]{Burgess:2005sb}%
  \BibitemOpen
  \bibfield  {author} {\bibinfo {author} {\bibfnamefont {C.~P.}\ \bibnamefont
  {Burgess}}, \bibinfo {author} {\bibfnamefont {R.}~\bibnamefont {Easther}},
  \bibinfo {author} {\bibfnamefont {A.}~\bibnamefont {Mazumdar}}, \bibinfo
  {author} {\bibfnamefont {D.~F.}\ \bibnamefont {Mota}},\ and\ \bibinfo
  {author} {\bibfnamefont {T.}~\bibnamefont {Multamaki}},\ }\bibfield  {title}
  {\bibinfo {title} {{Multiple inflation, cosmic string networks and the string
  landscape}},\ }\href {https://doi.org/10.1088/1126-6708/2005/05/067}
  {\bibfield  {journal} {\bibinfo  {journal} {JHEP}\ }\textbf {\bibinfo
  {volume} {05}},\ \bibinfo {pages} {067}},\ \Eprint
  {https://arxiv.org/abs/hep-th/0501125} {arXiv:hep-th/0501125 [hep-th]}
  \BibitemShut {NoStop}%
\bibitem [{\citenamefont {Wainwright}\ and\ \citenamefont
  {Profumo}(2009)}]{Wainwright:2009mq}%
  \BibitemOpen
  \bibfield  {author} {\bibinfo {author} {\bibfnamefont {C.}~\bibnamefont
  {Wainwright}}\ and\ \bibinfo {author} {\bibfnamefont {S.}~\bibnamefont
  {Profumo}},\ }\bibfield  {title} {\bibinfo {title} {{The Impact of a strongly
  first-order phase transition on the abundance of thermal relics}},\ }\href
  {https://doi.org/10.1103/PhysRevD.80.103517} {\bibfield  {journal} {\bibinfo
  {journal} {Phys. Rev. D}\ }\textbf {\bibinfo {volume} {80}},\ \bibinfo
  {pages} {103517} (\bibinfo {year} {2009})},\ \Eprint
  {https://arxiv.org/abs/0909.1317} {arXiv:0909.1317 [hep-ph]} \BibitemShut
  {NoStop}%
\bibitem [{\citenamefont {Davoudiasl}\ \emph {et~al.}(2016)\citenamefont
  {Davoudiasl}, \citenamefont {Hooper},\ and\ \citenamefont
  {McDermott}}]{Davoudiasl:2015vba}%
  \BibitemOpen
  \bibfield  {author} {\bibinfo {author} {\bibfnamefont {H.}~\bibnamefont
  {Davoudiasl}}, \bibinfo {author} {\bibfnamefont {D.}~\bibnamefont {Hooper}},\
  and\ \bibinfo {author} {\bibfnamefont {S.~D.}\ \bibnamefont {McDermott}},\
  }\bibfield  {title} {\bibinfo {title} {{Inflatable Dark Matter}},\ }\href
  {https://doi.org/10.1103/PhysRevLett.116.031303} {\bibfield  {journal}
  {\bibinfo  {journal} {Phys. Rev. Lett.}\ }\textbf {\bibinfo {volume} {116}},\
  \bibinfo {pages} {031303} (\bibinfo {year} {2016})},\ \Eprint
  {https://arxiv.org/abs/1507.08660} {arXiv:1507.08660 [hep-ph]} \BibitemShut
  {NoStop}%
\bibitem [{\citenamefont {Hoof}\ and\ \citenamefont
  {Jaeckel}(2017)}]{Hoof:2017ibo}%
  \BibitemOpen
  \bibfield  {author} {\bibinfo {author} {\bibfnamefont {S.}~\bibnamefont
  {Hoof}}\ and\ \bibinfo {author} {\bibfnamefont {J.}~\bibnamefont {Jaeckel}},\
  }\bibfield  {title} {\bibinfo {title} {{QCD axions and axionlike particles in
  a two-inflation scenario}},\ }\href
  {https://doi.org/10.1103/PhysRevD.96.115016} {\bibfield  {journal} {\bibinfo
  {journal} {Phys. Rev. D}\ }\textbf {\bibinfo {volume} {96}},\ \bibinfo
  {pages} {115016} (\bibinfo {year} {2017})},\ \Eprint
  {https://arxiv.org/abs/1709.01090} {arXiv:1709.01090 [hep-ph]} \BibitemShut
  {NoStop}%
\bibitem [{\citenamefont {Breitbach}\ \emph {et~al.}(2019)\citenamefont
  {Breitbach}, \citenamefont {Kopp}, \citenamefont {Madge}, \citenamefont
  {Opferkuch},\ and\ \citenamefont {Schwaller}}]{Breitbach:2018ddu}%
  \BibitemOpen
  \bibfield  {author} {\bibinfo {author} {\bibfnamefont {M.}~\bibnamefont
  {Breitbach}}, \bibinfo {author} {\bibfnamefont {J.}~\bibnamefont {Kopp}},
  \bibinfo {author} {\bibfnamefont {E.}~\bibnamefont {Madge}}, \bibinfo
  {author} {\bibfnamefont {T.}~\bibnamefont {Opferkuch}},\ and\ \bibinfo
  {author} {\bibfnamefont {P.}~\bibnamefont {Schwaller}},\ }\bibfield  {title}
  {\bibinfo {title} {{Dark, Cold, and Noisy: Constraining Secluded Hidden
  Sectors with Gravitational Waves}},\ }\href
  {https://doi.org/10.1088/1475-7516/2019/07/007} {\bibfield  {journal}
  {\bibinfo  {journal} {JCAP}\ }\textbf {\bibinfo {volume} {07}},\ \bibinfo
  {pages} {007}},\ \Eprint {https://arxiv.org/abs/1811.11175} {arXiv:1811.11175
  [hep-ph]} \BibitemShut {NoStop}%
\bibitem [{\citenamefont {Hambye}\ \emph {et~al.}(2018)\citenamefont {Hambye},
  \citenamefont {Strumia},\ and\ \citenamefont {Teresi}}]{Hambye:2018qjv}%
  \BibitemOpen
  \bibfield  {author} {\bibinfo {author} {\bibfnamefont {T.}~\bibnamefont
  {Hambye}}, \bibinfo {author} {\bibfnamefont {A.}~\bibnamefont {Strumia}},\
  and\ \bibinfo {author} {\bibfnamefont {D.}~\bibnamefont {Teresi}},\
  }\bibfield  {title} {\bibinfo {title} {{Super-cool Dark Matter}},\ }\href
  {https://doi.org/10.1007/JHEP08(2018)188} {\bibfield  {journal} {\bibinfo
  {journal} {JHEP}\ }\textbf {\bibinfo {volume} {08}},\ \bibinfo {pages}
  {188}},\ \Eprint {https://arxiv.org/abs/1805.01473} {arXiv:1805.01473
  [hep-ph]} \BibitemShut {NoStop}%
\bibitem [{\citenamefont {Hardy}\ and\ \citenamefont
  {Lasenby}(2017)}]{Hardy:2016kme}%
  \BibitemOpen
  \bibfield  {author} {\bibinfo {author} {\bibfnamefont {E.}~\bibnamefont
  {Hardy}}\ and\ \bibinfo {author} {\bibfnamefont {R.}~\bibnamefont
  {Lasenby}},\ }\bibfield  {title} {\bibinfo {title} {{Stellar cooling bounds
  on new light particles: plasma mixing effects}},\ }\href
  {https://doi.org/10.1007/JHEP02(2017)033} {\bibfield  {journal} {\bibinfo
  {journal} {JHEP}\ }\textbf {\bibinfo {volume} {02}},\ \bibinfo {pages}
  {033}},\ \Eprint {https://arxiv.org/abs/1611.05852} {arXiv:1611.05852
  [hep-ph]} \BibitemShut {NoStop}%
\bibitem [{\citenamefont {Knapen}\ \emph {et~al.}(2017)\citenamefont {Knapen},
  \citenamefont {Lin},\ and\ \citenamefont {Zurek}}]{Knapen:2017xzo}%
  \BibitemOpen
  \bibfield  {author} {\bibinfo {author} {\bibfnamefont {S.}~\bibnamefont
  {Knapen}}, \bibinfo {author} {\bibfnamefont {T.}~\bibnamefont {Lin}},\ and\
  \bibinfo {author} {\bibfnamefont {K.~M.}\ \bibnamefont {Zurek}},\ }\bibfield
  {title} {\bibinfo {title} {{Light Dark Matter: Models and Constraints}},\
  }\href {https://doi.org/10.1103/PhysRevD.96.115021} {\bibfield  {journal}
  {\bibinfo  {journal} {Phys. Rev. D}\ }\textbf {\bibinfo {volume} {96}},\
  \bibinfo {pages} {115021} (\bibinfo {year} {2017})},\ \Eprint
  {https://arxiv.org/abs/1709.07882} {arXiv:1709.07882 [hep-ph]} \BibitemShut
  {NoStop}%
\bibitem [{\citenamefont {Migdal}(1977)}]{Migdal:1977bq}%
  \BibitemOpen
  \bibfield  {author} {\bibinfo {author} {\bibfnamefont {A.~B.}\ \bibnamefont
  {Migdal}},\ }\href@noop {} {\emph {\bibinfo {title} {{Qualitative Methods in
  Quantum Theory}}}},\ Vol.~\bibinfo {volume} {48}\ (\bibinfo {year}
  {1977})\BibitemShut {NoStop}%
\bibitem [{\citenamefont {Landau}\ and\ \citenamefont
  {Lifshits}(1991)}]{Landau:1991wop}%
  \BibitemOpen
  \bibfield  {author} {\bibinfo {author} {\bibfnamefont {L.~D.}\ \bibnamefont
  {Landau}}\ and\ \bibinfo {author} {\bibfnamefont {E.}~\bibnamefont
  {Lifshits}},\ }\href@noop {} {\emph {\bibinfo {title} {{Quantum Mechanics}:
  {Non-Relativistic Theory}}}},\ \bibinfo {series} {Course of Theoretical
  Physics}, Vol.\ \bibinfo {volume} {v.3}\ (\bibinfo  {publisher}
  {Butterworth-Heinemann},\ \bibinfo {address} {Oxford},\ \bibinfo {year}
  {1991})\BibitemShut {NoStop}%
\bibitem [{\citenamefont {Ibe}\ \emph {et~al.}(2018{\natexlab{b}})\citenamefont
  {Ibe}, \citenamefont {Nakano}, \citenamefont {Shoji},\ and\ \citenamefont
  {Suzuki}}]{Ibe:2017yqa}%
  \BibitemOpen
  \bibfield  {author} {\bibinfo {author} {\bibfnamefont {M.}~\bibnamefont
  {Ibe}}, \bibinfo {author} {\bibfnamefont {W.}~\bibnamefont {Nakano}},
  \bibinfo {author} {\bibfnamefont {Y.}~\bibnamefont {Shoji}},\ and\ \bibinfo
  {author} {\bibfnamefont {K.}~\bibnamefont {Suzuki}},\ }\bibfield  {title}
  {\bibinfo {title} {{Migdal Effect in Dark Matter Direct Detection
  Experiments}},\ }\href {https://doi.org/10.1007/JHEP03(2018)194} {\bibfield
  {journal} {\bibinfo  {journal} {JHEP}\ }\textbf {\bibinfo {volume} {03}},\
  \bibinfo {pages} {194}},\ \Eprint {https://arxiv.org/abs/1707.07258}
  {arXiv:1707.07258 [hep-ph]} \BibitemShut {NoStop}%
\bibitem [{\citenamefont {Dolan}\ \emph {et~al.}(2018)\citenamefont {Dolan},
  \citenamefont {Kahlhoefer},\ and\ \citenamefont {McCabe}}]{Dolan:2017xbu}%
  \BibitemOpen
  \bibfield  {author} {\bibinfo {author} {\bibfnamefont {M.~J.}\ \bibnamefont
  {Dolan}}, \bibinfo {author} {\bibfnamefont {F.}~\bibnamefont {Kahlhoefer}},\
  and\ \bibinfo {author} {\bibfnamefont {C.}~\bibnamefont {McCabe}},\
  }\bibfield  {title} {\bibinfo {title} {{Directly detecting sub-GeV dark
  matter with electrons from nuclear scattering}},\ }\href
  {https://doi.org/10.1103/PhysRevLett.121.101801} {\bibfield  {journal}
  {\bibinfo  {journal} {Phys. Rev. Lett.}\ }\textbf {\bibinfo {volume} {121}},\
  \bibinfo {pages} {101801} (\bibinfo {year} {2018})},\ \Eprint
  {https://arxiv.org/abs/1711.09906} {arXiv:1711.09906 [hep-ph]} \BibitemShut
  {NoStop}%
\bibitem [{\citenamefont {Bernabei}\ \emph {et~al.}(2007)\citenamefont
  {Bernabei} \emph {et~al.}}]{Bernabei:2007jz}%
  \BibitemOpen
  \bibfield  {author} {\bibinfo {author} {\bibfnamefont {R.}~\bibnamefont
  {Bernabei}} \emph {et~al.},\ }\bibfield  {title} {\bibinfo {title} {{On
  electromagnetic contributions in WIMP quests}},\ }\href
  {https://doi.org/10.1142/S0217751X07037093} {\bibfield  {journal} {\bibinfo
  {journal} {Int. J. Mod. Phys. A}\ }\textbf {\bibinfo {volume} {22}},\
  \bibinfo {pages} {3155} (\bibinfo {year} {2007})},\ \Eprint
  {https://arxiv.org/abs/0706.1421} {arXiv:0706.1421 [astro-ph]} \BibitemShut
  {NoStop}%
\bibitem [{\citenamefont {Moustakidis}\ \emph {et~al.}(2005)\citenamefont
  {Moustakidis}, \citenamefont {Vergados},\ and\ \citenamefont
  {Ejiri}}]{Moustakidis:2005gx}%
  \BibitemOpen
  \bibfield  {author} {\bibinfo {author} {\bibfnamefont {C.}~\bibnamefont
  {Moustakidis}}, \bibinfo {author} {\bibfnamefont {J.}~\bibnamefont
  {Vergados}},\ and\ \bibinfo {author} {\bibfnamefont {H.}~\bibnamefont
  {Ejiri}},\ }\bibfield  {title} {\bibinfo {title} {{Direct dark matter
  detection by observing electrons produced in neutralino-nucleus
  collisions}},\ }\href {https://doi.org/10.1016/j.nuclphysb.2005.08.033}
  {\bibfield  {journal} {\bibinfo  {journal} {Nucl. Phys. B}\ }\textbf
  {\bibinfo {volume} {727}},\ \bibinfo {pages} {406} (\bibinfo {year}
  {2005})},\ \Eprint {https://arxiv.org/abs/hep-ph/0507123}
  {arXiv:hep-ph/0507123} \BibitemShut {NoStop}%
\bibitem [{\citenamefont {Ejiri}\ \emph {et~al.}(2006)\citenamefont {Ejiri},
  \citenamefont {Moustakidis},\ and\ \citenamefont {Vergados}}]{Ejiri:2005aj}%
  \BibitemOpen
  \bibfield  {author} {\bibinfo {author} {\bibfnamefont {H.}~\bibnamefont
  {Ejiri}}, \bibinfo {author} {\bibfnamefont {C.}~\bibnamefont {Moustakidis}},\
  and\ \bibinfo {author} {\bibfnamefont {J.}~\bibnamefont {Vergados}},\
  }\bibfield  {title} {\bibinfo {title} {{Dark matter search by exclusive
  studies of X-rays following WIMPs nuclear interactions}},\ }\href
  {https://doi.org/10.1016/j.physletb.2006.03.037} {\bibfield  {journal}
  {\bibinfo  {journal} {Phys. Lett. B}\ }\textbf {\bibinfo {volume} {639}},\
  \bibinfo {pages} {218} (\bibinfo {year} {2006})},\ \Eprint
  {https://arxiv.org/abs/hep-ph/0510042} {arXiv:hep-ph/0510042} \BibitemShut
  {NoStop}%
\bibitem [{\citenamefont {Baxter}\ \emph {et~al.}(2020)\citenamefont {Baxter},
  \citenamefont {Kahn},\ and\ \citenamefont {Krnjaic}}]{Baxter:2019pnz}%
  \BibitemOpen
  \bibfield  {author} {\bibinfo {author} {\bibfnamefont {D.}~\bibnamefont
  {Baxter}}, \bibinfo {author} {\bibfnamefont {Y.}~\bibnamefont {Kahn}},\ and\
  \bibinfo {author} {\bibfnamefont {G.}~\bibnamefont {Krnjaic}},\ }\bibfield
  {title} {\bibinfo {title} {{Electron Ionization via Dark Matter-Electron
  Scattering and the Migdal Effect}},\ }\href
  {https://doi.org/10.1103/PhysRevD.101.076014} {\bibfield  {journal} {\bibinfo
   {journal} {Phys. Rev. D}\ }\textbf {\bibinfo {volume} {101}},\ \bibinfo
  {pages} {076014} (\bibinfo {year} {2020})},\ \Eprint
  {https://arxiv.org/abs/1908.00012} {arXiv:1908.00012 [hep-ph]} \BibitemShut
  {NoStop}%
\bibitem [{\citenamefont {Knapen}\ \emph {et~al.}(2020)\citenamefont {Knapen},
  \citenamefont {Kozaczuk},\ and\ \citenamefont {Lin}}]{Knapen:2020aky}%
  \BibitemOpen
  \bibfield  {author} {\bibinfo {author} {\bibfnamefont {S.}~\bibnamefont
  {Knapen}}, \bibinfo {author} {\bibfnamefont {J.}~\bibnamefont {Kozaczuk}},\
  and\ \bibinfo {author} {\bibfnamefont {T.}~\bibnamefont {Lin}},\ }\bibfield
  {title} {\bibinfo {title} {{The Migdal effect in semiconductors}},\ }\Eprint
  {https://arxiv.org/abs/2011.09496} {arXiv:2011.09496 [hep-ph]}  (\bibinfo
  {year} {2020})\BibitemShut {NoStop}%
\bibitem [{\citenamefont {{Lotz}}(1967)}]{1967ApJS...14..207L}%
  \BibitemOpen
  \bibfield  {author} {\bibinfo {author} {\bibfnamefont {W.}~\bibnamefont
  {{Lotz}}},\ }\bibfield  {title} {\bibinfo {title} {{Electron-Impact
  Ionization Cross-Sections and Ionization Rate Coefficients for Atoms and
  Ions}},\ }\href {https://doi.org/10.1086/190154} {\bibfield  {journal}
  {\bibinfo  {journal} {Zeitschrift f{\"u}r Physik}\ }\textbf {\bibinfo
  {volume} {14}},\ \bibinfo {pages} {207} (\bibinfo {year} {1967})}\BibitemShut
  {NoStop}%
\bibitem [{\citenamefont {Massacrier}\ \emph {et~al.}(2011)\citenamefont
  {Massacrier}, \citenamefont {Potekhin},\ and\ \citenamefont
  {Chabrier}}]{Massacrier:2011kt}%
  \BibitemOpen
  \bibfield  {author} {\bibinfo {author} {\bibfnamefont {G.}~\bibnamefont
  {Massacrier}}, \bibinfo {author} {\bibfnamefont {A.}~\bibnamefont
  {Potekhin}},\ and\ \bibinfo {author} {\bibfnamefont {G.}~\bibnamefont
  {Chabrier}},\ }\bibfield  {title} {\bibinfo {title} {{Equation of state for
  partially ionized carbon and oxygen mixtures at high temperatures}},\ }\href
  {https://doi.org/10.1103/PhysRevE.84.056406} {\bibfield  {journal} {\bibinfo
  {journal} {Phys. Rev. E}\ }\textbf {\bibinfo {volume} {84}},\ \bibinfo
  {pages} {056406} (\bibinfo {year} {2011})},\ \Eprint
  {https://arxiv.org/abs/1111.0532} {arXiv:1111.0532 [physics.plasm-ph]}
  \BibitemShut {NoStop}%
\bibitem [{\citenamefont {Padmanabhan}(2000)}]{padmanabhan_2000}%
  \BibitemOpen
  \bibfield  {author} {\bibinfo {author} {\bibfnamefont {T.}~\bibnamefont
  {Padmanabhan}},\ }\href@noop {} {\emph {\bibinfo {title} {Theoretical
  Astrophysics}}},\ Vol.~\bibinfo {volume} {1}\ (\bibinfo  {publisher}
  {Cambridge University Press},\ \bibinfo {year} {2000})\BibitemShut {NoStop}%
\bibitem [{\citenamefont {Caughlan}\ and\ \citenamefont
  {Fowler}(1988)}]{Caughlan:1987qf}%
  \BibitemOpen
  \bibfield  {author} {\bibinfo {author} {\bibfnamefont {G.~R.}\ \bibnamefont
  {Caughlan}}\ and\ \bibinfo {author} {\bibfnamefont {W.~A.}\ \bibnamefont
  {Fowler}},\ }\bibfield  {title} {\bibinfo {title} {{Thermonuclear reaction
  rates. 5.}},\ }\href {https://doi.org/10.1016/0092-640X(88)90009-5}
  {\bibfield  {journal} {\bibinfo  {journal} {Atom. Data Nucl. Data Tabl.}\
  }\textbf {\bibinfo {volume} {40}},\ \bibinfo {pages} {283} (\bibinfo {year}
  {1988})}\BibitemShut {NoStop}%
\bibitem [{\citenamefont {Dziewonski}\ and\ \citenamefont
  {Anderson}(1981)}]{Dziewonski:1981xy}%
  \BibitemOpen
  \bibfield  {author} {\bibinfo {author} {\bibfnamefont {A.~M.}\ \bibnamefont
  {Dziewonski}}\ and\ \bibinfo {author} {\bibfnamefont {D.~L.}\ \bibnamefont
  {Anderson}},\ }\bibfield  {title} {\bibinfo {title} {{Preliminary reference
  earth model}},\ }\href {https://doi.org/10.1016/0031-9201(81)90046-7}
  {\bibfield  {journal} {\bibinfo  {journal} {Phys. Earth Planet. Interiors}\
  }\textbf {\bibinfo {volume} {25}},\ \bibinfo {pages} {297} (\bibinfo {year}
  {1981})}\BibitemShut {NoStop}%
\bibitem [{\citenamefont {Clarke}\ and\ \citenamefont
  {Washington}(1924)}]{clarke1924composition}%
  \BibitemOpen
  \bibfield  {author} {\bibinfo {author} {\bibfnamefont {F.}~\bibnamefont
  {Clarke}}\ and\ \bibinfo {author} {\bibfnamefont {H.}~\bibnamefont
  {Washington}},\ }\href {https://books.google.ca/books?id=Ht9WAAAAMAAJ} {\emph
  {\bibinfo {title} {The Composition of the Earth's Crust}}},\ \bibinfo
  {series} {Geological Survey professional paper}\ No.\ \bibinfo {number} {nos.
  126-127}\ (\bibinfo  {publisher} {U.S. Government Printing Office},\ \bibinfo
  {year} {1924})\BibitemShut {NoStop}%
\bibitem [{\citenamefont {Wang}\ \emph {et~al.}(2018)\citenamefont {Wang},
  \citenamefont {Lineweaver},\ and\ \citenamefont {Ireland}}]{WANG2018460}%
  \BibitemOpen
  \bibfield  {author} {\bibinfo {author} {\bibfnamefont {H.~S.}\ \bibnamefont
  {Wang}}, \bibinfo {author} {\bibfnamefont {C.~H.}\ \bibnamefont
  {Lineweaver}},\ and\ \bibinfo {author} {\bibfnamefont {T.~R.}\ \bibnamefont
  {Ireland}},\ }\bibfield  {title} {\bibinfo {title} {The elemental abundances
  (with uncertainties) of the most earth-like planet},\ }\href
  {https://doi.org/10.1016/j.icarus.2017.08.024} {\bibfield  {journal}
  {\bibinfo  {journal} {Icarus}\ }\textbf {\bibinfo {volume} {299}},\ \bibinfo
  {pages} {460 } (\bibinfo {year} {2018})}\BibitemShut {NoStop}%
\bibitem [{\citenamefont {Morgan}\ and\ \citenamefont
  {Anders}(1980)}]{Morgan6973}%
  \BibitemOpen
  \bibfield  {author} {\bibinfo {author} {\bibfnamefont {J.~W.}\ \bibnamefont
  {Morgan}}\ and\ \bibinfo {author} {\bibfnamefont {E.}~\bibnamefont
  {Anders}},\ }\bibfield  {title} {\bibinfo {title} {Chemical composition of
  earth, venus, and mercury},\ }\href {https://doi.org/10.1073/pnas.77.12.6973}
  {\bibfield  {journal} {\bibinfo  {journal} {Proceedings of the National
  Academy of Sciences}\ }\textbf {\bibinfo {volume} {77}},\ \bibinfo {pages}
  {6973} (\bibinfo {year} {1980})}\BibitemShut {NoStop}%
\bibitem [{\citenamefont {McDonough}(2003)}]{MCDONOUGH2003547}%
  \BibitemOpen
  \bibfield  {author} {\bibinfo {author} {\bibfnamefont {W.}~\bibnamefont
  {McDonough}},\ }\bibfield  {title} {\bibinfo {title} {2.15 - compositional
  model for the earth's core},\ }in\ \href
  {https://doi.org/10.1016/B0-08-043751-6/02015-6} {\emph {\bibinfo {booktitle}
  {Treatise on Geochemistry}}},\ \bibinfo {editor} {edited by\ \bibinfo
  {editor} {\bibfnamefont {H.~D.}\ \bibnamefont {Holland}}\ and\ \bibinfo
  {editor} {\bibfnamefont {K.~K.}\ \bibnamefont {Turekian}}}\ (\bibinfo
  {publisher} {Pergamon},\ \bibinfo {address} {Oxford},\ \bibinfo {year}
  {2003})\ pp.\ \bibinfo {pages} {547 -- 568}\BibitemShut {NoStop}%
\bibitem [{\citenamefont {Johnston}\ \emph {et~al.}(1974)\citenamefont
  {Johnston}, \citenamefont {McGetchin},\ and\ \citenamefont
  {Toksöz}}]{johnston1974}%
  \BibitemOpen
  \bibfield  {author} {\bibinfo {author} {\bibfnamefont {D.~H.}\ \bibnamefont
  {Johnston}}, \bibinfo {author} {\bibfnamefont {T.~R.}\ \bibnamefont
  {McGetchin}},\ and\ \bibinfo {author} {\bibfnamefont {M.~N.}\ \bibnamefont
  {Toksöz}},\ }\bibfield  {title} {\bibinfo {title} {The thermal state and
  internal structure of mars},\ }\href
  {https://doi.org/10.1029/JB079i026p03959} {\bibfield  {journal} {\bibinfo
  {journal} {Journal of Geophysical Research (1896-1977)}\ }\textbf {\bibinfo
  {volume} {79}},\ \bibinfo {pages} {3959} (\bibinfo {year}
  {1974})}\BibitemShut {NoStop}%
\bibitem [{\citenamefont {Bramante}\ \emph {et~al.}(2018)\citenamefont
  {Bramante}, \citenamefont {Broerman}, \citenamefont {Lang},\ and\
  \citenamefont {Raj}}]{Bramante:2018qbc}%
  \BibitemOpen
  \bibfield  {author} {\bibinfo {author} {\bibfnamefont {J.}~\bibnamefont
  {Bramante}}, \bibinfo {author} {\bibfnamefont {B.}~\bibnamefont {Broerman}},
  \bibinfo {author} {\bibfnamefont {R.~F.}\ \bibnamefont {Lang}},\ and\
  \bibinfo {author} {\bibfnamefont {N.}~\bibnamefont {Raj}},\ }\bibfield
  {title} {\bibinfo {title} {{Saturated Overburden Scattering and the
  Multiscatter Frontier: Discovering Dark Matter at the Planck Mass and
  Beyond}},\ }\href {https://doi.org/10.1103/PhysRevD.98.083516} {\bibfield
  {journal} {\bibinfo  {journal} {Phys. Rev.}\ }\textbf {\bibinfo {volume}
  {D98}},\ \bibinfo {pages} {083516} (\bibinfo {year} {2018})},\ \Eprint
  {https://arxiv.org/abs/1803.08044} {arXiv:1803.08044 [hep-ph]} \BibitemShut
  {NoStop}%
\bibitem [{\citenamefont {Aartsen}\ \emph {et~al.}(2017)\citenamefont {Aartsen}
  \emph {et~al.}}]{Aartsen:2016nxy}%
  \BibitemOpen
  \bibfield  {author} {\bibinfo {author} {\bibfnamefont {M.}~\bibnamefont
  {Aartsen}} \emph {et~al.} (\bibinfo {collaboration} {IceCube}),\ }\bibfield
  {title} {\bibinfo {title} {{The IceCube Neutrino Observatory: Instrumentation
  and Online Systems}},\ }\href
  {https://doi.org/10.1088/1748-0221/12/03/P03012} {\bibfield  {journal}
  {\bibinfo  {journal} {JINST}\ }\textbf {\bibinfo {volume} {12}}\bibfield
  {number} {\bibinfo  {number} { (03)},\ \bibinfo {pages} {P03012}},\ }\Eprint
  {https://arxiv.org/abs/1612.05093} {arXiv:1612.05093 [astro-ph.IM]}
  \BibitemShut {NoStop}%
\bibitem [{\citenamefont {Bellini}\ \emph {et~al.}(2014)\citenamefont {Bellini}
  \emph {et~al.}}]{Bellini:2013lnn}%
  \BibitemOpen
  \bibfield  {author} {\bibinfo {author} {\bibfnamefont {G.}~\bibnamefont
  {Bellini}} \emph {et~al.} (\bibinfo {collaboration} {Borexino}),\ }\bibfield
  {title} {\bibinfo {title} {{Final results of Borexino Phase-I on low energy
  solar neutrino spectroscopy}},\ }\href
  {https://doi.org/10.1103/PhysRevD.89.112007} {\bibfield  {journal} {\bibinfo
  {journal} {Phys. Rev. D}\ }\textbf {\bibinfo {volume} {89}},\ \bibinfo
  {pages} {112007} (\bibinfo {year} {2014})},\ \Eprint
  {https://arxiv.org/abs/1308.0443} {arXiv:1308.0443 [hep-ex]} \BibitemShut
  {NoStop}%
\bibitem [{\citenamefont {Andringa}\ \emph {et~al.}(2016)\citenamefont
  {Andringa} \emph {et~al.}}]{Andringa:2015tza}%
  \BibitemOpen
  \bibfield  {author} {\bibinfo {author} {\bibfnamefont {S.}~\bibnamefont
  {Andringa}} \emph {et~al.} (\bibinfo {collaboration} {SNO+}),\ }\bibfield
  {title} {\bibinfo {title} {{Current Status and Future Prospects of the SNO+
  Experiment}},\ }\href {https://doi.org/10.1155/2016/6194250} {\bibfield
  {journal} {\bibinfo  {journal} {Adv. High Energy Phys.}\ }\textbf {\bibinfo
  {volume} {2016}},\ \bibinfo {pages} {6194250} (\bibinfo {year} {2016})},\
  \Eprint {https://arxiv.org/abs/1508.05759} {arXiv:1508.05759
  [physics.ins-det]} \BibitemShut {NoStop}%
\bibitem [{\citenamefont {Joglekar}\ \emph
  {et~al.}(2020{\natexlab{a}})\citenamefont {Joglekar}, \citenamefont {Raj},
  \citenamefont {Tanedo},\ and\ \citenamefont {Yu}}]{Joglekar:2019vzy}%
  \BibitemOpen
  \bibfield  {author} {\bibinfo {author} {\bibfnamefont {A.}~\bibnamefont
  {Joglekar}}, \bibinfo {author} {\bibfnamefont {N.}~\bibnamefont {Raj}},
  \bibinfo {author} {\bibfnamefont {P.}~\bibnamefont {Tanedo}},\ and\ \bibinfo
  {author} {\bibfnamefont {H.-B.}\ \bibnamefont {Yu}},\ }\bibfield  {title}
  {\bibinfo {title} {{Relativistic capture of dark matter by electrons in
  neutron stars}},\ }\href {https://doi.org/10.1016/j.physletb.2020.135767}
  {\bibfield  {journal} {\bibinfo  {journal} {Phys. Lett.}\ }\textbf {\bibinfo
  {volume} {B}},\ \bibinfo {pages} {135767} (\bibinfo {year}
  {2020}{\natexlab{a}})},\ \Eprint {https://arxiv.org/abs/1911.13293}
  {arXiv:1911.13293 [hep-ph]} \BibitemShut {NoStop}%
\bibitem [{\citenamefont {Joglekar}\ \emph
  {et~al.}(2020{\natexlab{b}})\citenamefont {Joglekar}, \citenamefont {Raj},
  \citenamefont {Tanedo},\ and\ \citenamefont {Yu}}]{Joglekar:2020liw}%
  \BibitemOpen
  \bibfield  {author} {\bibinfo {author} {\bibfnamefont {A.}~\bibnamefont
  {Joglekar}}, \bibinfo {author} {\bibfnamefont {N.}~\bibnamefont {Raj}},
  \bibinfo {author} {\bibfnamefont {P.}~\bibnamefont {Tanedo}},\ and\ \bibinfo
  {author} {\bibfnamefont {H.-B.}\ \bibnamefont {Yu}},\ }\bibfield  {title}
  {\bibinfo {title} {{Dark kinetic heating of neutron stars from contact
  interactions with relativistic targets}},\ }\href
  {https://doi.org/10.1103/PhysRevD.102.123002} {\bibfield  {journal} {\bibinfo
   {journal} {Phys. Rev. D}\ }\textbf {\bibinfo {volume} {102}},\ \bibinfo
  {pages} {123002} (\bibinfo {year} {2020}{\natexlab{b}})},\ \Eprint
  {https://arxiv.org/abs/2004.09539} {arXiv:2004.09539 [hep-ph]} \BibitemShut
  {NoStop}%
\bibitem [{\citenamefont {Acevedo}\ \emph {et~al.}(2022)\citenamefont
  {Acevedo}, \citenamefont {Bramante},\ and\ \citenamefont
  {Goodman}}]{Acevedo:2021kly}%
  \BibitemOpen
  \bibfield  {author} {\bibinfo {author} {\bibfnamefont {J.~F.}\ \bibnamefont
  {Acevedo}}, \bibinfo {author} {\bibfnamefont {J.}~\bibnamefont {Bramante}},\
  and\ \bibinfo {author} {\bibfnamefont {A.}~\bibnamefont {Goodman}},\
  }\bibfield  {title} {\bibinfo {title} {{Accelerating composite dark matter
  discovery with nuclear recoils and the Migdal effect}},\ }\href
  {https://doi.org/10.1103/PhysRevD.105.023012} {\bibfield  {journal} {\bibinfo
   {journal} {Phys. Rev. D}\ }\textbf {\bibinfo {volume} {105}},\ \bibinfo
  {pages} {023012} (\bibinfo {year} {2022})},\ \Eprint
  {https://arxiv.org/abs/2108.10889} {arXiv:2108.10889 [hep-ph]} \BibitemShut
  {NoStop}%
\bibitem [{\citenamefont {Bramante}\ \emph {et~al.}(2020)\citenamefont
  {Bramante}, \citenamefont {Buchanan}, \citenamefont {Goodman},\ and\
  \citenamefont {Lodhi}}]{Bramante:2019fhi}%
  \BibitemOpen
  \bibfield  {author} {\bibinfo {author} {\bibfnamefont {J.}~\bibnamefont
  {Bramante}}, \bibinfo {author} {\bibfnamefont {A.}~\bibnamefont {Buchanan}},
  \bibinfo {author} {\bibfnamefont {A.}~\bibnamefont {Goodman}},\ and\ \bibinfo
  {author} {\bibfnamefont {E.}~\bibnamefont {Lodhi}},\ }\bibfield  {title}
  {\bibinfo {title} {{Terrestrial and Martian Heat Flow Limits on Dark
  Matter}},\ }\href {https://doi.org/10.1103/PhysRevD.101.043001} {\bibfield
  {journal} {\bibinfo  {journal} {Phys. Rev.}\ }\textbf {\bibinfo {volume}
  {D101}},\ \bibinfo {pages} {043001} (\bibinfo {year} {2020})},\ \Eprint
  {https://arxiv.org/abs/1909.11683} {arXiv:1909.11683 [hep-ph]} \BibitemShut
  {NoStop}%
\bibitem [{\citenamefont {Acevedo}\ \emph {et~al.}(2021)\citenamefont
  {Acevedo}, \citenamefont {Bramante}, \citenamefont {Goodman}, \citenamefont
  {Kopp},\ and\ \citenamefont {Opferkuch}}]{Acevedo:2020gro}%
  \BibitemOpen
  \bibfield  {author} {\bibinfo {author} {\bibfnamefont {J.~F.}\ \bibnamefont
  {Acevedo}}, \bibinfo {author} {\bibfnamefont {J.}~\bibnamefont {Bramante}},
  \bibinfo {author} {\bibfnamefont {A.}~\bibnamefont {Goodman}}, \bibinfo
  {author} {\bibfnamefont {J.}~\bibnamefont {Kopp}},\ and\ \bibinfo {author}
  {\bibfnamefont {T.}~\bibnamefont {Opferkuch}},\ }\bibfield  {title} {\bibinfo
  {title} {{Dark Matter, Destroyer of Worlds: Neutrino, Thermal, and
  Existential Signatures from Black Holes in the Sun and Earth}},\ }\href
  {https://doi.org/10.1088/1475-7516/2021/04/026} {\bibfield  {journal}
  {\bibinfo  {journal} {JCAP}\ }\textbf {\bibinfo {volume} {04}},\ \bibinfo
  {pages} {026}},\ \Eprint {https://arxiv.org/abs/2012.09176} {arXiv:2012.09176
  [hep-ph]} \BibitemShut {NoStop}%
\bibitem [{\citenamefont {Mack}\ \emph {et~al.}(2007)\citenamefont {Mack},
  \citenamefont {Beacom},\ and\ \citenamefont {Bertone}}]{Mack:2007xj}%
  \BibitemOpen
  \bibfield  {author} {\bibinfo {author} {\bibfnamefont {G.~D.}\ \bibnamefont
  {Mack}}, \bibinfo {author} {\bibfnamefont {J.~F.}\ \bibnamefont {Beacom}},\
  and\ \bibinfo {author} {\bibfnamefont {G.}~\bibnamefont {Bertone}},\
  }\bibfield  {title} {\bibinfo {title} {{Towards Closing the Window on
  Strongly Interacting Dark Matter: Far-Reaching Constraints from Earth's Heat
  Flow}},\ }\href {https://doi.org/10.1103/PhysRevD.76.043523} {\bibfield
  {journal} {\bibinfo  {journal} {Phys. Rev. D}\ }\textbf {\bibinfo {volume}
  {76}},\ \bibinfo {pages} {043523} (\bibinfo {year} {2007})},\ \Eprint
  {https://arxiv.org/abs/0705.4298} {arXiv:0705.4298 [astro-ph]} \BibitemShut
  {NoStop}%
\bibitem [{\citenamefont {Bramante}(2015)}]{Bramante:2015cua}%
  \BibitemOpen
  \bibfield  {author} {\bibinfo {author} {\bibfnamefont {J.}~\bibnamefont
  {Bramante}},\ }\bibfield  {title} {\bibinfo {title} {{Dark matter ignition of
  type Ia supernovae}},\ }\href
  {https://doi.org/10.1103/PhysRevLett.115.141301} {\bibfield  {journal}
  {\bibinfo  {journal} {Phys. Rev. Lett.}\ }\textbf {\bibinfo {volume} {115}},\
  \bibinfo {pages} {141301} (\bibinfo {year} {2015})},\ \Eprint
  {https://arxiv.org/abs/1505.07464} {arXiv:1505.07464 [hep-ph]} \BibitemShut
  {NoStop}%
\bibitem [{\citenamefont {Graham}\ \emph {et~al.}(2015)\citenamefont {Graham},
  \citenamefont {Rajendran},\ and\ \citenamefont {Varela}}]{Graham:2015apa}%
  \BibitemOpen
  \bibfield  {author} {\bibinfo {author} {\bibfnamefont {P.~W.}\ \bibnamefont
  {Graham}}, \bibinfo {author} {\bibfnamefont {S.}~\bibnamefont {Rajendran}},\
  and\ \bibinfo {author} {\bibfnamefont {J.}~\bibnamefont {Varela}},\
  }\bibfield  {title} {\bibinfo {title} {{Dark Matter Triggers of
  Supernovae}},\ }\href {https://doi.org/10.1103/PhysRevD.92.063007} {\bibfield
   {journal} {\bibinfo  {journal} {Phys. Rev. D}\ }\textbf {\bibinfo {volume}
  {92}},\ \bibinfo {pages} {063007} (\bibinfo {year} {2015})},\ \Eprint
  {https://arxiv.org/abs/1505.04444} {arXiv:1505.04444 [hep-ph]} \BibitemShut
  {NoStop}%
\bibitem [{\citenamefont {Graham}\ \emph {et~al.}(2018)\citenamefont {Graham},
  \citenamefont {Janish}, \citenamefont {Narayan}, \citenamefont {Rajendran},\
  and\ \citenamefont {Riggins}}]{Graham:2018efk}%
  \BibitemOpen
  \bibfield  {author} {\bibinfo {author} {\bibfnamefont {P.~W.}\ \bibnamefont
  {Graham}}, \bibinfo {author} {\bibfnamefont {R.}~\bibnamefont {Janish}},
  \bibinfo {author} {\bibfnamefont {V.}~\bibnamefont {Narayan}}, \bibinfo
  {author} {\bibfnamefont {S.}~\bibnamefont {Rajendran}},\ and\ \bibinfo
  {author} {\bibfnamefont {P.}~\bibnamefont {Riggins}},\ }\bibfield  {title}
  {\bibinfo {title} {{White Dwarfs as Dark Matter Detectors}},\ }\href
  {https://doi.org/10.1103/PhysRevD.98.115027} {\bibfield  {journal} {\bibinfo
  {journal} {Phys. Rev. D}\ }\textbf {\bibinfo {volume} {98}},\ \bibinfo
  {pages} {115027} (\bibinfo {year} {2018})},\ \Eprint
  {https://arxiv.org/abs/1805.07381} {arXiv:1805.07381 [hep-ph]} \BibitemShut
  {NoStop}%
\bibitem [{\citenamefont {Acevedo}\ and\ \citenamefont
  {Bramante}(2019)}]{Acevedo:2019gre}%
  \BibitemOpen
  \bibfield  {author} {\bibinfo {author} {\bibfnamefont {J.~F.}\ \bibnamefont
  {Acevedo}}\ and\ \bibinfo {author} {\bibfnamefont {J.}~\bibnamefont
  {Bramante}},\ }\bibfield  {title} {\bibinfo {title} {{Supernovae Sparked By
  Dark Matter in White Dwarfs}},\ }\href
  {https://doi.org/10.1103/PhysRevD.100.043020} {\bibfield  {journal} {\bibinfo
   {journal} {Phys. Rev.}\ }\textbf {\bibinfo {volume} {D100}},\ \bibinfo
  {pages} {043020} (\bibinfo {year} {2019})},\ \Eprint
  {https://arxiv.org/abs/1904.11993} {arXiv:1904.11993 [hep-ph]} \BibitemShut
  {NoStop}%
\bibitem [{\citenamefont {Janish}\ \emph {et~al.}(2019)\citenamefont {Janish},
  \citenamefont {Narayan},\ and\ \citenamefont {Riggins}}]{Janish:2019nkk}%
  \BibitemOpen
  \bibfield  {author} {\bibinfo {author} {\bibfnamefont {R.}~\bibnamefont
  {Janish}}, \bibinfo {author} {\bibfnamefont {V.}~\bibnamefont {Narayan}},\
  and\ \bibinfo {author} {\bibfnamefont {P.}~\bibnamefont {Riggins}},\
  }\bibfield  {title} {\bibinfo {title} {{Type Ia supernovae from dark matter
  core collapse}},\ }\href {https://doi.org/10.1103/PhysRevD.100.035008}
  {\bibfield  {journal} {\bibinfo  {journal} {Phys. Rev. D}\ }\textbf {\bibinfo
  {volume} {100}},\ \bibinfo {pages} {035008} (\bibinfo {year} {2019})},\
  \Eprint {https://arxiv.org/abs/1905.00395} {arXiv:1905.00395 [hep-ph]}
  \BibitemShut {NoStop}%
\bibitem [{\citenamefont {Fedderke}\ \emph {et~al.}(2020)\citenamefont
  {Fedderke}, \citenamefont {Graham},\ and\ \citenamefont
  {Rajendran}}]{Fedderke:2019jur}%
  \BibitemOpen
  \bibfield  {author} {\bibinfo {author} {\bibfnamefont {M.~A.}\ \bibnamefont
  {Fedderke}}, \bibinfo {author} {\bibfnamefont {P.~W.}\ \bibnamefont
  {Graham}},\ and\ \bibinfo {author} {\bibfnamefont {S.}~\bibnamefont
  {Rajendran}},\ }\bibfield  {title} {\bibinfo {title} {{White dwarf bounds on
  charged massive particles}},\ }\href
  {https://doi.org/10.1103/PhysRevD.101.115021} {\bibfield  {journal} {\bibinfo
   {journal} {Phys. Rev. D}\ }\textbf {\bibinfo {volume} {101}},\ \bibinfo
  {pages} {115021} (\bibinfo {year} {2020})},\ \Eprint
  {https://arxiv.org/abs/1911.08883} {arXiv:1911.08883 [hep-ph]} \BibitemShut
  {NoStop}%
\bibitem [{\citenamefont {{Timmes}}\ and\ \citenamefont
  {{Woosley}}(1992)}]{1992Timmes}%
  \BibitemOpen
  \bibfield  {author} {\bibinfo {author} {\bibfnamefont {F.~X.}\ \bibnamefont
  {{Timmes}}}\ and\ \bibinfo {author} {\bibfnamefont {S.~E.}\ \bibnamefont
  {{Woosley}}},\ }\bibfield  {title} {\bibinfo {title} {{The conductive
  propagation of nuclear flames. I - Degenerate C $+$ O and O $+$ NE $+$ MG
  white dwarfs}},\ }\href {https://doi.org/10.1086/171746} {\bibfield
  {journal} {\bibinfo  {journal} {Astrophys. J.}\ }\textbf {\bibinfo {volume}
  {396}},\ \bibinfo {pages} {649} (\bibinfo {year} {1992})}\BibitemShut
  {NoStop}%
\bibitem [{\citenamefont {Potekhin}\ \emph {et~al.}(1999)\citenamefont
  {Potekhin}, \citenamefont {Baiko}, \citenamefont {Haensel},\ and\
  \citenamefont {Yakovlev}}]{Potekhin:1999yv}%
  \BibitemOpen
  \bibfield  {author} {\bibinfo {author} {\bibfnamefont {A.}~\bibnamefont
  {Potekhin}}, \bibinfo {author} {\bibfnamefont {D.}~\bibnamefont {Baiko}},
  \bibinfo {author} {\bibfnamefont {P.}~\bibnamefont {Haensel}},\ and\ \bibinfo
  {author} {\bibfnamefont {D.}~\bibnamefont {Yakovlev}},\ }\bibfield  {title}
  {\bibinfo {title} {{Transport properties of degenerate electrons in neutron
  star envelopes and white dwarf cores}},\ }\href@noop {} {\bibfield  {journal}
  {\bibinfo  {journal} {Astron. Astrophys.}\ }\textbf {\bibinfo {volume}
  {346}},\ \bibinfo {pages} {345} (\bibinfo {year} {1999})},\ \Eprint
  {https://arxiv.org/abs/astro-ph/9903127} {arXiv:astro-ph/9903127}
  \BibitemShut {NoStop}%
\bibitem [{\citenamefont {Kippenhahn}\ \emph {et~al.}(2012)\citenamefont
  {Kippenhahn}, \citenamefont {Weigert},\ and\ \citenamefont
  {Weiss}}]{Kippenhahn:1994wva}%
  \BibitemOpen
  \bibfield  {author} {\bibinfo {author} {\bibfnamefont {R.}~\bibnamefont
  {Kippenhahn}}, \bibinfo {author} {\bibfnamefont {A.}~\bibnamefont
  {Weigert}},\ and\ \bibinfo {author} {\bibfnamefont {A.}~\bibnamefont
  {Weiss}},\ }\href {https://doi.org/10.1007/978-3-642-30304-3} {\emph
  {\bibinfo {title} {{Stellar structure and evolution}}}},\ Vol.\ \bibinfo
  {volume} {9783642303043}\ (\bibinfo  {publisher} {Springer},\ \bibinfo {year}
  {2012})\BibitemShut {NoStop}%
\bibitem [{\citenamefont {Gasques}\ \emph {et~al.}(2005)\citenamefont
  {Gasques}, \citenamefont {Afanasjev}, \citenamefont {Aguilera}, \citenamefont
  {Beard}, \citenamefont {Chamon}, \citenamefont {Ring}, \citenamefont
  {Wiescher},\ and\ \citenamefont {Yakovlev}}]{Gasques:2005ar}%
  \BibitemOpen
  \bibfield  {author} {\bibinfo {author} {\bibfnamefont {L.}~\bibnamefont
  {Gasques}}, \bibinfo {author} {\bibfnamefont {A.}~\bibnamefont {Afanasjev}},
  \bibinfo {author} {\bibfnamefont {E.}~\bibnamefont {Aguilera}}, \bibinfo
  {author} {\bibfnamefont {M.}~\bibnamefont {Beard}}, \bibinfo {author}
  {\bibfnamefont {L.}~\bibnamefont {Chamon}}, \bibinfo {author} {\bibfnamefont
  {P.}~\bibnamefont {Ring}}, \bibinfo {author} {\bibfnamefont {M.}~\bibnamefont
  {Wiescher}},\ and\ \bibinfo {author} {\bibfnamefont {D.}~\bibnamefont
  {Yakovlev}},\ }\bibfield  {title} {\bibinfo {title} {{Nuclear fusion in dense
  matter: Reaction rate and carbon burning}},\ }\href
  {https://doi.org/10.1103/PhysRevC.72.025806} {\bibfield  {journal} {\bibinfo
  {journal} {Phys. Rev. C}\ }\textbf {\bibinfo {volume} {72}},\ \bibinfo
  {pages} {025806} (\bibinfo {year} {2005})},\ \Eprint
  {https://arxiv.org/abs/astro-ph/0506386} {arXiv:astro-ph/0506386}
  \BibitemShut {NoStop}%
\bibitem [{\citenamefont {Descouvemont}\ \emph {et~al.}(2004)\citenamefont
  {Descouvemont}, \citenamefont {Adahchour}, \citenamefont {Angulo},
  \citenamefont {Coc},\ and\ \citenamefont
  {Vangioni-Flam}}]{Descouvemont:2004cw}%
  \BibitemOpen
  \bibfield  {author} {\bibinfo {author} {\bibfnamefont {P.}~\bibnamefont
  {Descouvemont}}, \bibinfo {author} {\bibfnamefont {A.}~\bibnamefont
  {Adahchour}}, \bibinfo {author} {\bibfnamefont {C.}~\bibnamefont {Angulo}},
  \bibinfo {author} {\bibfnamefont {A.}~\bibnamefont {Coc}},\ and\ \bibinfo
  {author} {\bibfnamefont {E.}~\bibnamefont {Vangioni-Flam}},\ }\bibfield
  {title} {\bibinfo {title} {{Compilation and R-matrix analysis of Big Bang
  nuclear reaction rates}},\ }\href {https://doi.org/10.1016/j.adt.2004.08.001}
  {\bibfield  {journal} {\bibinfo  {journal} {Atom. Data Nucl. Data Tabl.}\
  }\textbf {\bibinfo {volume} {88}},\ \bibinfo {pages} {203} (\bibinfo {year}
  {2004})},\ \Eprint {https://arxiv.org/abs/astro-ph/0407101}
  {arXiv:astro-ph/0407101} \BibitemShut {NoStop}%
\bibitem [{\citenamefont {Ando}\ \emph {et~al.}(2006)\citenamefont {Ando},
  \citenamefont {Cyburt}, \citenamefont {Hong},\ and\ \citenamefont
  {Hyun}}]{Ando:2005cz}%
  \BibitemOpen
  \bibfield  {author} {\bibinfo {author} {\bibfnamefont {S.}~\bibnamefont
  {Ando}}, \bibinfo {author} {\bibfnamefont {R.}~\bibnamefont {Cyburt}},
  \bibinfo {author} {\bibfnamefont {S.}~\bibnamefont {Hong}},\ and\ \bibinfo
  {author} {\bibfnamefont {C.}~\bibnamefont {Hyun}},\ }\bibfield  {title}
  {\bibinfo {title} {{Radiative neutron capture on a proton at BBN energies}},\
  }\href {https://doi.org/10.1103/PhysRevC.74.025809} {\bibfield  {journal}
  {\bibinfo  {journal} {Phys. Rev. C}\ }\textbf {\bibinfo {volume} {74}},\
  \bibinfo {pages} {025809} (\bibinfo {year} {2006})},\ \Eprint
  {https://arxiv.org/abs/nucl-th/0511074} {arXiv:nucl-th/0511074} \BibitemShut
  {NoStop}%
\bibitem [{\citenamefont {Jedamzik}\ and\ \citenamefont
  {Pospelov}(2009)}]{Jedamzik:2009uy}%
  \BibitemOpen
  \bibfield  {author} {\bibinfo {author} {\bibfnamefont {K.}~\bibnamefont
  {Jedamzik}}\ and\ \bibinfo {author} {\bibfnamefont {M.}~\bibnamefont
  {Pospelov}},\ }\bibfield  {title} {\bibinfo {title} {{Big Bang
  Nucleosynthesis and Particle Dark Matter}},\ }\href
  {https://doi.org/10.1088/1367-2630/11/10/105028} {\bibfield  {journal}
  {\bibinfo  {journal} {New J. Phys.}\ }\textbf {\bibinfo {volume} {11}},\
  \bibinfo {pages} {105028} (\bibinfo {year} {2009})},\ \Eprint
  {https://arxiv.org/abs/0906.2087} {arXiv:0906.2087 [hep-ph]} \BibitemShut
  {NoStop}%
\bibitem [{\citenamefont {Greiner}(1990)}]{Greiner:1990tz}%
  \BibitemOpen
  \bibfield  {author} {\bibinfo {author} {\bibfnamefont {W.}~\bibnamefont
  {Greiner}},\ }\href@noop {} {\emph {\bibinfo {title} {{Relativistic quantum
  mechanics: Wave equations}}}}\ (\bibinfo {year} {1990})\BibitemShut {NoStop}%
\bibitem [{\citenamefont {Manzur}\ \emph {et~al.}(2010)\citenamefont {Manzur},
  \citenamefont {Curioni}, \citenamefont {Kastens}, \citenamefont {McKinsey},
  \citenamefont {Ni},\ and\ \citenamefont {Wongjirad}}]{Manzur:2009hp}%
  \BibitemOpen
  \bibfield  {author} {\bibinfo {author} {\bibfnamefont {A.}~\bibnamefont
  {Manzur}}, \bibinfo {author} {\bibfnamefont {A.}~\bibnamefont {Curioni}},
  \bibinfo {author} {\bibfnamefont {L.}~\bibnamefont {Kastens}}, \bibinfo
  {author} {\bibfnamefont {D.~N.}\ \bibnamefont {McKinsey}}, \bibinfo {author}
  {\bibfnamefont {K.}~\bibnamefont {Ni}},\ and\ \bibinfo {author}
  {\bibfnamefont {T.}~\bibnamefont {Wongjirad}},\ }\bibfield  {title} {\bibinfo
  {title} {{Scintillation efficiency and ionization yield of liquid xenon for
  mono-energetic nuclear recoils down to 4 keV}},\ }\href
  {https://doi.org/10.1103/PhysRevC.81.025808} {\bibfield  {journal} {\bibinfo
  {journal} {Phys. Rev. C}\ }\textbf {\bibinfo {volume} {81}},\ \bibinfo
  {pages} {025808} (\bibinfo {year} {2010})},\ \Eprint
  {https://arxiv.org/abs/0909.1063} {arXiv:0909.1063 [physics.ins-det]}
  \BibitemShut {NoStop}%
\bibitem [{\citenamefont {Kimura}\ \emph {et~al.}(2019)\citenamefont {Kimura},
  \citenamefont {Tanaka}, \citenamefont {Washimi},\ and\ \citenamefont
  {Yorita}}]{Kimura:2019rdg}%
  \BibitemOpen
  \bibfield  {author} {\bibinfo {author} {\bibfnamefont {M.}~\bibnamefont
  {Kimura}}, \bibinfo {author} {\bibfnamefont {M.}~\bibnamefont {Tanaka}},
  \bibinfo {author} {\bibfnamefont {T.}~\bibnamefont {Washimi}},\ and\ \bibinfo
  {author} {\bibfnamefont {K.}~\bibnamefont {Yorita}},\ }\bibfield  {title}
  {\bibinfo {title} {{Measurement of the scintillation efficiency for nuclear
  recoils in liquid argon under electric fields up to 3 kV/cm}},\ }\href
  {https://doi.org/10.1103/PhysRevD.100.032002} {\bibfield  {journal} {\bibinfo
   {journal} {Phys. Rev. D}\ }\textbf {\bibinfo {volume} {100}},\ \bibinfo
  {pages} {032002} (\bibinfo {year} {2019})},\ \Eprint
  {https://arxiv.org/abs/1902.01501} {arXiv:1902.01501 [physics.ins-det]}
  \BibitemShut {NoStop}%
\bibitem [{\citenamefont {McCabe}(2017)}]{McCabe:2017rln}%
  \BibitemOpen
  \bibfield  {author} {\bibinfo {author} {\bibfnamefont {C.}~\bibnamefont
  {McCabe}},\ }\bibfield  {title} {\bibinfo {title} {{New constraints and
  discovery potential of sub-GeV dark matter with xenon detectors}},\ }\href
  {https://doi.org/10.1103/PhysRevD.96.043010} {\bibfield  {journal} {\bibinfo
  {journal} {Phys. Rev. D}\ }\textbf {\bibinfo {volume} {96}},\ \bibinfo
  {pages} {043010} (\bibinfo {year} {2017})},\ \Eprint
  {https://arxiv.org/abs/1702.04730} {arXiv:1702.04730 [hep-ph]} \BibitemShut
  {NoStop}%
\end{thebibliography}%

\end{document}